\documentclass[pre,twocolumn,showpacs]{revtex4}

\usepackage{graphicx}

\usepackage{amsmath,amssymb}
\usepackage{bm}
\usepackage{cancel}

\usepackage{color}

\begin{document}
\title{ Soliton-like attractor for blood vessel tip density in angiogenesis} 
\author{ L.\ L.\ Bonilla, M.\ Carretero, and F.\ Terragni },
\affiliation{$^1$G. Mill\'an Institute, Fluid Dynamics, Nanoscience and Industrial Mathematics, Universidad Carlos III de Madrid, 28911 Legan\'es, Spain}

\date{December 16th, 2016}
\begin{abstract} 
Recently, numerical simulations of a stochastic model have shown that the density of vessel tips in tumor induced angiogenesis adopts a soliton-like profile [Sci. Rep. 6, 31296 (2016)]. In this work, we derive and solve the equations for the soliton collective coordinates that indicate how the soliton adapts its shape and velocity to varying chemotaxis and diffusion. The vessel tip density can be reconstructed from the soliton formulas. While the stochastic model exhibits large fluctuations, we show that the location of the maximum vessel tip density for different replicas follows closely the soliton peak position calculated either by ensemble averages or by solving an alternative deterministic description of the density. The simple soliton collective coordinate equations may also be used to ascertain the response of the vessel network to changes in the parameters and thus to control it. 
\end{abstract}

\maketitle

\section{Introduction}
\label{sec:intro}
The growth of blood vessels is a complex multiscale process called angiogenesis that is the basis of organ growth and repair in healthy conditions and also of pathological developments such as cancerous tumors \cite{fol74,car05,fig08,car11}. Cells in an incipient tumor located in tissue experience lack of oxygen and nutrients, and stimulate production of vessel endothelial growth factor that, in turn, induces growth of blood vessels (angiogenesis) from a nearby primary vessel in the tumor direction \cite{fol74,car05}. Blood brings oxygen and nutrients that foster tumor growth. In angiogenesis, events happening in cellular and subcellular scales unchain endothelial cell motion and proliferation, build millimeter scale blood sprouts and networks thereof \cite{car11,CT05,GG05,fru07}. Angiogenesis imbalance contributes to numerous malignant, inflammatory, ischaemic, infectious, and immune disorders \cite{car05}. For these reasons, immense human and material resources are devoted to understanding and controlling angiogenesis. Theoretical efforts based on angiogenesis models go hand in hand with experiments \cite{lio77,sto91,cha93,cha95,and98,ton01,lev01,pla03,man04,sun05a,sun05,ste06,bau07,cap09,jac10,das10,swa11,sci11,sci13,cot14,dej14,ben14,bon14,hec15}. Models range from very simple to extraordinarily complex and often try to illuminate some particular mechanism; see the review \cite{hec15}. Realistic microscopic models involve postulating mechanisms and a large number of parameters that cannot be directly estimated from experiments, but they often yield qualitative predictions that can be tested. An important challenge is to extract mesoscopic and macroscopic descriptions of angiogenesis from the diverse microscopic models. 

Early angiogenesis macroscopic models consisted of reaction-diffusion equations for densities of cell and chemicals (growth factors, fibronectin, etc.) \cite{lio77,cha93,cha95}. These models do not allow to treat the growth and evolution of individual blood vessels. Later models focused on the evolution of the cells at the tip of a vessel sprout. The ten or so cells at a vessel tip are highly motile and do not proliferate. They follow chemotactic and haptotactic clues as they advance toward hypoxic regions that experience lack of oxygen. These cells are followed by proliferating stalk cells that build a capillary in their wake. Thus {\em tip cell models} are based on the motion of single particles representing the tip cells and their trajectories constitute the advancing blood vessels \cite{sto91,and98,pla03,man04,cap09,bon14,hec15,ter16}. More realistic and necessarily more complex models illuminate  tip and stalk cell dynamics, the motion of tip and stalk cells on the extracellular matrix outside blood vessels, blood circulation in newly formed vessels, and so on \cite{bau07,jac10,ben14,hec15}. 

In recent work \cite{bon14,ter16}, we have been trying to  bridge the gap between microscopic descriptions of early stage tumor induced angiogenesis that require large numerical simulations and macroscopic descriptions that are amenable to a more thorough theoretical study. We consider a simple tip cell model in which tip stochastic extension is driven by the gradient of growth factors (chemotaxis), there is a random branching of tips and tips join with existing blood vessels (anastomosis). We have derived a deterministic description for the density of vessel tips consisting of an integrodifferential equation for the tip density coupled to a reaction-diffusion equation for the tumor angiogenic factor (TAF, which comprises vessel endothelial and other growth factors) \cite{bon14,ter16}. The stochastic model can be made more realistic by adding equations characterizing haptotaxis, the influence of other chemicals or drugs, etc. While cell densities can be extracted from numerical simulations of microscopic models, our equation for the tip density \cite{bon14} incorporates tip branching and anastomosis as derived from a stochastic model \cite{ter16}, not postulated ad hoc. It turns out that the tip density soon forms a moving lump that advances towards the tumor. The longitudinal section of the stable lump (that we may term {\em angiton}) is approximately given by a moving soliton-like wave \cite{bon16}. This wave is an exact 1D solution of a reduced equation for the marginal tip density on the whole real line that has constant chemotactic force and no diffusion. It appears by differentiating a domain-wall solution (topological soliton) connecting two spatially homogeneous states. Numerical evidence shows that it is asymptotically stable \cite{bon16}. Technically speaking, it is not known whether two soliton-like waves in the angiogenesis model equations emerge unchanged from collisions except for a phase shift. Therefore we do not claim that angiogenesis soliton-like lump profiles are true solitons. However stable soliton-like waves are central to the arguments of the present paper and, by an abuse of language, we will call them solitons. In this, we follow extended usage in the physical literature in which other stable waves such as ``topological solitons'' \cite{MS04} or ``diffusive solitons'' \cite{rem99} are called simply solitons despite not emerging unscathed from collisions \cite{MS04,rem99}. The soliton shape and velocity depend on two collective coordinates. The vessel tip density approaches the soliton solution after an initial formation stage. After its formation and until the vessels are close to the tumor, the tip density is described by the soliton and the solution of its two collective coordinate equations.

In this paper, we deduce the equations for the angiogenesis soliton and its collective coordinates, solve the latter numerically and reconstruct the marginal tip density from the soliton formula. Then we show that it agrees with both the solution of the deterministic description and with the ensemble average of the tip density as extracted from the stochastic process. Although the fluctuations are large, we give numerical evidence that the position of the soliton peak is very close to that of the maximum of the marginal tip density for different replicas or realizations of the stochastic process. This implies that the simple description based on the soliton may give useful information about single replicas of the angiogenesis process. While our simple model needs to be completed to discuss control of angiogenesis, we show how changing a single parameter results in seemingly arresting the process.

The rest of the paper is as follows. We recall the stochastic model of \cite{bon14} and its deterministic description \cite{ter16} in Section \ref{sec:model}. By a Chapman-Enskog method, we derive a reduced equation for the marginal tip density in Section \ref{sec:reduced}. By neglecting diffusion and considering constant coefficients in the resulting equation, we find in Section \ref{sec:soliton} an analytical expression for the soliton of the marginal tip density \cite{bon16}. Section \ref{sec:cc} contains a derivation of the differential equations for the two collective coordinates of the soliton. The coefficients appearing in these equations contain spatial averages of the TAF density. In Section \ref{sec:numerical}, we explain how to calculate the coefficients in the collective coordinate equations, solve them numerically, reconstruct the soliton and, through it, the marginal vessel tip density. We compare it with direct solutions of the deterministic description and ensemble averages of the stochastic process. Although realizations of the stochastic angiogenic process provide very different looking vessel networks, we also show that the maximum of the marginal density for each realization follows closely the soliton peak. Section \ref{sec:conclusions} contains our conclusions and the Appendices are devoted to technical matters.  

\section{Model} \label{sec:model}
Early stages of angiogenesis are described by a simple stochastic model in \cite{bon14,ter16}. It consists of a system of Langevin equations for the extension of vessel tips, a tip branching process and tip annihilation (anastomosis) when they merge with existing vessels. A tip $i$ is born at a random time $T^i$ from a moving tip (we ignore branching from mature vessels) and disappears at a later random time $\Theta^i$, either by reaching the tumor or by anastomosis. At time $T^i$, the velocity of the newly created tip $i$ is selected out of a normal distribution,
\begin{eqnarray}
\delta_{\sigma_v}(\mathbf{v}-\mathbf{v}_0)= \frac{e^{-|\mathbf{v}-\mathbf{v}_0|^2/\sigma_v^2}}{\pi\sigma_v^2},\label{eq1}
\end{eqnarray}
with mean $\mathbf{v}_0$ and a narrow variance $\sigma_v^2$. In addition, the probability that a tip branches from one of the existing ones during an infinitesimal time interval $(t, t + dt]$ is taken proportional to $\sum_{i=1}^{N(t)}\alpha(C(t,\mathbf{X}^i(t)))dt$, where $C(t,\mathbf{x})$ is the TAF concentration and 
\begin{eqnarray}
\alpha(C)=\alpha_1\frac{C}{C_R+C},\quad C_R>0,\,\, \alpha_1>0,\label{eq2}
\end{eqnarray}
in which $C_R$ is a reference concentration. The change per unit time of the number of tips in boxes $d\mathbf{x}$ and $d\mathbf{v}$ about $\mathbf{x}$ and $\mathbf{v}$ is
\begin{eqnarray}\nonumber
&&\sum_{i=1}^{N(t)}\alpha(C(t,\mathbf{X}^i(t)))\, \delta_{\sigma_v}(\mathbf{v}^i(t)-\mathbf{v}_0)=\int_{d\mathbf{x}}\int_{d\mathbf{v}}\alpha(C(t,\mathbf{x)})\\
&&\times \delta_{\sigma_v}(\mathbf{v}-\mathbf{v}_0)
\sum_{i=1}^{N(t)}\delta(\mathbf{x}-\mathbf{X}^i(t))\delta(\mathbf{v}-\mathbf{v}^i(t)) d\mathbf{x} d\mathbf{v}. \label{eq3}
\end{eqnarray}

The Langevin equations for tip extensions are
\begin{eqnarray} 
&&d\mathbf{X}^i(t)=\mathbf{v}^i(t)\, dt,\nonumber\\
&&d\mathbf{v}^i(t)= \left[- k\, \mathbf{v}^i(t)+\mathbf{F}\!\left(C(t,\mathbf{X}^i(t))\right)\!\right]\! dt + \sigma\, d\mathbf{W}^i(t), \label{eq4}
\end{eqnarray}
where $\mathbf{X}^i(t)$ and $\mathbf{v}^i(t)$ are the tip position and velocity of tip $i$ at time $t$, $\mathbf{W}^i(t)$ are independent identically distributed (i.i.d.) standard Brownian motions, and $k$ (friction coefficient) and $\sigma$ are positive parameters. At each time $t$ there are $N(t)$ active tips. The chemotactic force is 
\begin{eqnarray}
\mathbf{F}(C)&=& \frac{d_1}{(1+\gamma_1C)^q}\nabla_x C,  \label{eq5}
\end{eqnarray}
where $d_1$, $\gamma_1$, and $q$ are positive parameters. The TAF concentration solves
\begin{eqnarray} 
\frac{\partial}{\partial t}C(t,\mathbf{x})\!&\!=\!& \! d_2 \Delta_x C(t,\mathbf{x})
-\eta C(t,\mathbf{x})\nonumber \\&\times&
\!\left| \sum_{i=1}^{N(t)} \mathbf{v}^i(t)\delta_{\sigma_x}(\mathbf{x}-\mathbf{X}^i(t))\right|\!.\label{eq6}
\end{eqnarray}
Here $d_2$ (diffusivity) and $\eta$ are positive parameters, whereas $\delta_{\sigma_x}(\mathbf{x})$ is a regularized smooth delta function (e.g., a Gaussian with variances $l_x^2$ and $l_y^2$ proportional to $\sigma_x^2$ along the $x$ and $y$ directions, respectively) that becomes $\delta(\mathbf{x})$ in the limit as $\sigma_x\to 0$.

There is a counterpart to the stochastic model for the densities of vessel tips and the vessel tip flux, defined as ensemble averages over a sufficient number $\mathcal{N}$ of replicas (realizations) $\omega$ of the stochastic process:
\begin{eqnarray}
p_{\mathcal{N}}\!(t,\mathbf{x},\mathbf{v})\!&=&\!\frac{1}{\mathcal{N}}\sum_{\omega=1}^\mathcal{N}\sum_{i=1}^{N(t,\omega)}\delta_{\sigma_x}(\mathbf{x}-\mathbf{X}^i(t,\omega))\nonumber\\&\times&
\delta_{\sigma_v}(\mathbf{v}-\mathbf{v}^i(t,\omega)),\label{eq7}\\
\tilde{p}_{\mathcal N}(t,\mathbf{x})\!\!&=&\!\frac{1}{\mathcal{N}}\sum_{\omega=1}^\mathcal{N}\sum_{i=1}^{N(t,\omega)}\delta_{\sigma_x}(\mathbf{x}-\mathbf{X}^i(t,\omega)), \label{eq8}\\
\mathbf{j}_{\mathcal N}(t,\mathbf{x})\!\!&=&\!\frac{1}{\mathcal{N}}\!\sum_{\omega=1}^\mathcal{N}\!\sum_{i=1}^{N(t,\omega)}\!\!\mathbf{v}^i(t,\omega)\delta_{\sigma_x}(\mathbf{x}-\mathbf{X}^i(t,\omega)).\label{eq9}
\end{eqnarray}
As $\mathcal{N}\to\infty$, these ensemble averages tend to the tip density $p(t,\mathbf{x},\mathbf{v})$, the marginal tip density $\tilde{p}(t,\mathbf{x})$, and the tip flux $\mathbf{j}(t,\mathbf{x})$, respectively. In \cite{ter16} it is shown that the angiogenesis model has a deterministic description based on the following equation for the density of vessel tips, $p(t,\mathbf{x},\mathbf{v})$, 
\begin{eqnarray}
&&\frac{\partial}{\partial t} p(t,\mathbf{x},\mathbf{v})=\alpha(C(t,\mathbf{x}))\,
 p(t,\mathbf{x},\mathbf{v})\delta_{v}(\mathbf{v}-\mathbf{v}_0)\nonumber\\
&& - \gamma\, p(t,\mathbf{x},\mathbf{v}) \int_0^t \tilde{p}(s,\mathbf{x})\, ds  - \mathbf{v}\cdot \nabla_x   p(t,\mathbf{x},\mathbf{v}) \nonumber\\
&& - \nabla_v \cdot [(\mathbf{F}(C(t,\mathbf{x}))-k\mathbf{v}) p(t,\mathbf{x},\mathbf{v})]\nonumber\\ 
&&+ \frac{\sigma^2}{2} \Delta_{v} p(t,\mathbf{x},\mathbf{v}),
\label{eq10}\\
&& \tilde{p}(t,\mathbf{x})=\int p(t,\mathbf{x},\mathbf{v}')\, d \mathbf{v'}. \label{eq11}
\end{eqnarray}
The TAF equation \eqref{eq6} becomes
\begin{eqnarray} 
\frac{\partial}{\partial t}C(t,\mathbf{x})=d_2 \Delta_x C(t,\mathbf{x})- \eta\, C(t,\mathbf{x})\!\left| \mathbf{j}(t,\mathbf{x})\right|\!,\label{eq12}
\end{eqnarray}
where $\mathbf{j}(t,\mathbf{x})$ is the current density (flux) vector at any point $\mathbf{x}$ and any time $t\geq 0$,
\begin{equation}
\mathbf{j}(t,\mathbf{x})= \int \mathbf{v}'
p(t,\mathbf{x},\mathbf{v}')\, d \mathbf{v'}. \label{eq13}
\end{equation}
Alternatively, if $N(t)$ becomes very large (which is precluded by anastomosis), the same deterministic description can be derived by using the law of large numbers \cite{bon14}.
\begin{table}[ht]
\begin{center}\begin{tabular}{ccccccc}
 \hline
$\mathbf{x}$& $\mathbf{v}$ & $t$ &$C$& $p$ &$\tilde{p}$&$\mathbf{j}$\\
$L$ & $\tilde{v}_0$ & $\frac{L}{\tilde{v}_0}$ & $C_R$ & $\frac{1}{\tilde{v}_0^2L^2}$& $\frac{1}{L^2}$& $\frac{\tilde{v}_0}{L^2}$\\
mm&$\mu$m/hr & hr& mol/m$^2$&$10^{21}\frac{\mbox{s$^2$}}{\mbox{m$^4$}}$ & $10^{5}$m$^{-2}$&m$^{-1}$s$^{-1}$\\
$2$& 40 & 50 & $10^{-16}$ & 2.025 & 2.5 & 0.0028 \\ 
\hline
\end{tabular}
\end{center}
\caption{Units for nondimensionalizing the model equations. }
\label{table1}
\end{table}

The deterministic description consisting of Equations \eqref{eq10} and \eqref{eq12} is well posed, as it has been proved to have unique smooth solutions \cite{car16}. After nondimensionalization as in Table \ref{table1} \cite{bon14,ter16}, \eqref{eq10} and \eqref{eq12} become
\begin{eqnarray}
&&\frac{\partial}{\partial t} p(t,\mathbf{x},\mathbf{v})=
\frac{A\, C(t,\mathbf{x})}{1+C(t,\mathbf{x})}\,
 p(t,\mathbf{x},\mathbf{v})\delta_{v}(\mathbf{v}-\mathbf{v}_0)\nonumber\\
&& - \Gamma p(t,\mathbf{x},\mathbf{v}) \int_0^t \int p(s,\mathbf{x},\mathbf{v}')\, d \mathbf{v'} ds  - \mathbf{v}\cdot \nabla_x   p(t,\mathbf{x},\mathbf{v}) \nonumber\\
&& - \nabla_v \cdot\left[\!\left(\frac{\delta\,\nabla_x C(t,\mathbf{x})}{[1+\Gamma_1C(t,\mathbf{x})]^q}-\beta\mathbf{v}\right) p(t,\mathbf{x},\mathbf{v})
\right]\nonumber\\ 
&&+ \frac{\beta}{2} \Delta_{v} p(t,\mathbf{x},\mathbf{v}),
\label{eq14}\\
&&\frac{\partial}{\partial t}C(t,\mathbf{x})=\kappa \Delta_x C(t,\mathbf{x})- \chi\, C(t,\mathbf{x})\!\left| \mathbf{j}(t,\mathbf{x})\right|\!,\label{eq15}
\end{eqnarray}
respectively. The dimensionless parameters are defined in Table \ref{table2} and the boundary conditions to solve \eqref{eq14}-\eqref{eq15} are listed in Appendix \ref{app2}.
\begin{table}[ht]
\begin{center}\begin{tabular}{cccccccc}
 \hline
$\delta$ & $\beta$ &$A$& $\Gamma$& $\Gamma_1$ &$\kappa$&$\chi$&$\sigma_v$\\
 $\frac{d_1C_R}{\tilde{v}_0^2}$ & $\frac{kL}{\tilde{v}_0}$ & $\frac{\alpha_1L}{\tilde{v}_0^3}$ & $\frac{\gamma}{\tilde{v}_0^2}$&$\gamma_1C_R$& $\frac{d_2}{\tilde{v}_0 L}$& $\frac{\eta}{L}$&-\\
1.5 & 5.88 & $22.42$ & 0.145 & 1& $0.0045$ & 0.002&0.08 \\
 \hline
\end{tabular}
\end{center}
\caption{Dimensionless parameters. } 
\label{table2}
\end{table}

\section{Reduced equation for the marginal tip density}\label{sec:reduced}
We can obtain a simpler equation for the marginal vessel tip density \eqref{eq11} provided the overall tip density approaches rapidly a local equilibrium which is a displaced Maxwellian:
\begin{eqnarray}
p^{(0)}(t,\mathbf{x},\mathbf{v})=\frac{1}{\pi}e^{-|\mathbf{v}-\mathbf{v}_0|^2}\tilde{p}(t,\mathbf{x}).\label{eq16}
\end{eqnarray}
The source terms in \eqref{eq14} (two first terms on its right hand side) select velocities on a small neighborhood of $\mathbf{v}_0$, as such velocities are the only ones for which the birth term proportional to $\alpha(C)\delta_v(\mathbf{v}-\mathbf{v}_0)$, cf Eq.\ \eqref{eq1}, can compensate the anastomosis death term. To derive the simpler equation for $\tilde{p}$, we use the Chapman-Enskog method \cite{BT10}. We first rewrite (\ref{eq14}) as
\begin{eqnarray}
\mathcal{L}p &\equiv &\beta\,\nabla_v\cdot\left(\frac{1}{2}\nabla_vp+(\mathbf{v}-\mathbf{v}_0)p\right)\nonumber\\
&=& \epsilon\left[\frac{\partial p}{\partial t} +\beta\, (\mathbf{F}-\mathbf{v}_0)\cdot\nabla_v p +\mathbf{v}\nabla_xp \right.\nonumber\\&-&\left.
\alpha  p\,\delta_{v}(\mathbf{v}-\mathbf{v}_0) +\Gamma p \int_0^t \tilde{p}(s,\mathbf{x})\, ds\right]\!, \label{eq17}\\
\alpha&=&\frac{A\, C}{1+C},\label{eq18}\\
\mathbf{F}&=& \frac{\delta}{\beta}\,\frac{\nabla_x C(t,\mathbf{x})}{[1+\Gamma_1C(t,\mathbf{x})]^q}.\label{eq19}
\end{eqnarray}
We have included a scaling parameter $\epsilon$ in the right hand side of (\ref{eq17}), as we will consider that it is small compared to the left hand side. After the computations that follow, we will restore $\epsilon=1$. Note that (\ref{eq16}) satisfies 
\begin{eqnarray}
\mathcal{L}p^{(0)} =0,\label{eq20}
\end{eqnarray}
i.e., (\ref{eq17}) with $\epsilon=0$. We now assume that the terms on the right hand side of (\ref{eq17}) are small compared to those on its left hand side (formally, $\epsilon\ll 1$) and that we can expand $p$ in the asymptotic series
\begin{eqnarray}
p=p^{(0)} +\epsilon p^{(1)} + \epsilon^2 p^{(2)} +\ldots .\label{eq21}
\end{eqnarray}
Inserting this into (\ref{eq11}), we find
\begin{eqnarray}
\int p^{(j)} d\mathbf{v} =0,\quad j=1,2,\ldots. \label{eq22}
\end{eqnarray}
We assume now that 
\begin{eqnarray}
\frac{\partial\tilde{p}}{\partial t}=\mathcal{F}^{(0)} +\epsilon\mathcal{F}^{(1)}+\ldots,\label{eq23}
\end{eqnarray}
where the $\mathcal{F}^{(j)}$ should be determined by solvability conditions to be derived below. Inserting (\ref{eq21}) and (\ref{eq23}) in (\ref{eq17}) and equating like powers of $\epsilon$ in the result, we obtain the hierarchy of equations (\ref{eq20}) and
\begin{eqnarray}
&&\mathcal{L}p^{(1)}\! =\frac{e^{- V^2}}{\pi}\!\left[\mathcal{F}^{(0)} +\mathbf{v}\cdot\nabla_x\tilde{p} - 2\beta\mathbf{V}\!\cdot\! (\mathbf{F}-\mathbf{v}_0)\tilde{p} \right.\nonumber\\&&\quad\left.
- \alpha\tilde{p} \delta_{v}(\mathbf{V})+ \Gamma\tilde{p} \int_0^t \tilde{p}(s,\mathbf{x})\, ds\right]\!,\label{eq24}\\
&&\mathcal{L}p^{(2)}\! =\frac{e^{-V^2}}{\pi}\mathcal{F}^{(1)} +\mathbf{v}\cdot\nabla_xp^{(1)}\nonumber\\&&\quad
-2\beta\mathbf{V}\!\cdot\! (\mathbf{F}-\mathbf{v}_0)p^{(1)}- \alpha p^{(1)} \delta_{v}(\mathbf{V}) \nonumber\\&& \quad
+ \Gamma p^{(1)} \int_0^t \tilde{p}(s,\mathbf{x})\, ds,\label{eq25}
\end{eqnarray}
etc. Here $\mathbf{V}=\mathbf{v}-\mathbf{v}_0$ and $V=|\mathbf{V}|$. For these equations to have bounded solutions, we need to impose the conditions
\begin{eqnarray}
\int \mathcal{L}p^{(j)} d\mathbf{v} =0,\quad j=1,2,\ldots, \label{eq26}
\end{eqnarray}
as the adjoint problem $\mathcal{L}^\dagger v=0$ has constant solutions. For (\ref{eq24}), this condition yields
\begin{eqnarray}
\mathcal{F}^{(0)} =\frac{\alpha}{\pi}\tilde{p} -\mathbf{v}_0\cdot\nabla_x\tilde{p}- \Gamma\tilde{p} \int_0^t \tilde{p}(s,\mathbf{x})\, ds, \label{eq27}
\end{eqnarray}
which, inserted back in (\ref{eq24}), produces the equation
\begin{eqnarray}
\mathcal{L}p^{(1)} &=&\frac{e^{- V^2}}{\pi}\!\left\{\alpha\!\left[\frac{1}{\pi}  -\delta_{v}(\mathbf{V})\right]\!\tilde{p}  \right.\nonumber\\ &+&\left. 
\mathbf{V}\cdot\!\left[\nabla_x\tilde{p}-2\beta(\mathbf{F}-\mathbf{v}_0)\tilde{p}\right]\right\}.\label{eq28}
\end{eqnarray}
The solution of (\ref{eq28}) that satisfies (\ref{eq22}) is
\begin{eqnarray}
p^{(1)} =-\frac{e^{- V^2}}{\pi}\mathbf{V}\!\cdot\!\left[\nabla_x\tilde{p}-2\beta(\mathbf{F}-\mathbf{v}_0)\tilde{p}\right] \nonumber\\
+\frac{\alpha\tilde{p}}{2\pi^2}e^{-V^2} \!\left[ \int_0^\infty e^{-t}\ln t \, dt -\ln V^2\right]\!. \label{eq29}
\end{eqnarray}
Insertion of (\ref{eq29}) into the solvability condition (\ref{eq26}) for $j=2$ produces
\begin{eqnarray}
\mathcal{F}^{(1)} &=&\frac{1}{2\beta}\Delta_x\tilde{p}+\nabla_x\cdot\!\left[\left(\mathbf{v}_0- \mathbf{F}\right)\tilde{p}\right] \nonumber\\&+&
\frac{\alpha^2\tilde{p}}{2\pi^2\beta (1+\sigma_v^2)}\ln\!\left(1+\frac{1}{\sigma_v^2}\right)\!.\label{eq30}
\end{eqnarray}
We now substitute (\ref{eq27}) and (\ref{eq30}) in (\ref{eq23}) and recall $\epsilon=1$, thereby finding the Smoluchowski-type equation
\begin{eqnarray}
\frac{\partial\tilde{p}}{\partial t}+\nabla_x\cdot(\mathbf{F}\tilde{p})-\frac{1}{2\beta}\Delta_x\tilde{p}=\mu\,\tilde{p}\nonumber\\
-\Gamma\tilde{p}\int_0^t\tilde{p}(s,\mathbf{x})\, ds, \label{eq31}\\
\mu=\frac{\alpha}{\pi}\left[1+\frac{\alpha}{2\pi\beta(1+\sigma_v^2)}\ln\!\left(1+\frac{1}{\sigma_v^2}\right)\!\right]\!.\label{eq32}
\end{eqnarray}
Note that the convective terms in \eqref{eq31} correspond to having ignored inertia in the Langevin equation \eqref{eq4}, which then becomes $d\mathbf{X}^i(t)=(\mathbf{F}/k)\, dt + (\sigma/k)\, d\mathbf{W}^i (t)$. Our perturbation procedure just renormalizes the birth term $\alpha(C)$ in \eqref{eq14} or \eqref{eq17}.

The flux (\ref{eq13}) in the reaction-diffusion equation (\ref{eq15}) is $\mathbf{j}(t,\mathbf{x})\approx \mathbf{v}_0\tilde{p}(t,\mathbf{x})$, so that (\ref{eq15}) becomes
\begin{eqnarray} 
\frac{\partial}{\partial t}C(t,\mathbf{x})=\kappa \Delta_x C(t,\mathbf{x})- \chi\, C(t,\mathbf{x})\,\tilde{p}(t,\mathbf{x}),\label{eq33}
\end{eqnarray}
because $|\mathbf{v}_0|=1$ in our nondimensional units.

The boundary conditions for (\ref{eq31}) are: (i) $\tilde{p}(t,\mathbf{x})$ known at $x=1$ and equal to its instantaneous value there; and (ii) known flux $j_0$ at $x=0$ \cite{bon14}. The boundary condition (i) is a free boundary condition that avoids modeling explicitly the tumor instead of the more appropriate absorbing boundary condition $\tilde{p}=0$ at the tumor. In condition (ii), the flux can be approximated as
\begin{eqnarray}\nonumber
\int (\mathbf{v}_0+\mathbf{V})p(t,\mathbf{x},\mathbf{v})\, d\mathbf{V}&=& \mathbf{v}_0\tilde{p}+\int \mathbf{V} p^{(1)}d\mathbf{V} \\&=& 
\mathbf{F}\tilde{p}- \frac{1}{2\beta} \nabla_x \tilde{p}. \nonumber
\end{eqnarray}
At $x=0$, the x-component of $\mathbf{F}$ is zero and therefore the boundary condition for $\tilde{p}$ becomes $-\frac{1}{2\beta}\frac{\partial\tilde{p}}{\partial x}= j_0$, i.e.,
\begin{eqnarray}
\left.-\frac{1}{2\beta}\frac{\partial\tilde{p}}{\partial x}\right|_{x=0}= v_0\mu\,\tilde{p}\, \theta(\tau-t),  \label{eq34}
\end{eqnarray}
in which $\theta(t)=1$ if $t>0$ and $\theta(t)=0$ otherwise is the unit step function. In \eqref{eq34}, we have renormalized the birth rate coefficient $\alpha$ to $\mu$ in harmony with the change in birth rate when going from the equation for the vessel tip density \eqref{eq14} to \eqref{eq31} for the marginal vessel tip density; see \eqref{b6} in Appendix \ref{app2}. In \cite{bon14,ter16} and in the numerical calculations of this paper, $\tau=\infty$. 

\section{Soliton}
\label{sec:soliton}
We now find an approximate soliton solution of (\ref{eq31}) following \cite{bon16}. Firstly, let define
\begin{eqnarray}
\rho(t,\mathbf{x}) = \int_0^t\tilde{p}(s,\mathbf{x})\, ds,   \label{eq35}
\end{eqnarray}
and ignore diffusion in (\ref{eq31}), which then becomes
\begin{eqnarray}
&&\frac{\partial^2\rho}{\partial t^2}+\nabla_x\cdot\!\left(\mathbf{F}\frac{\partial\rho}{\partial t}\right)\! =\mu\frac{\partial\rho}{\partial t}-\Gamma\rho\frac{\partial\rho}{\partial t}. \label{eq36}
\end{eqnarray}
The coefficients $\kappa$ and $\chi$ in (\ref{eq33}) are very small \cite{bon14} and therefore the TAF concentration varies very slowly compared with the marginal tip density. We will also assume that the initial TAF concentration varies on a larger spatial scale than the soliton size and that the TAF gradient is directed on the $x$ axis, which constitutes a good approximation \cite{bon14}. Then $\mathbf{F}$ and $\mu$ are almost constant and we will seek a solution of the form
\begin{eqnarray}
\rho(t,\mathbf{x}) = \rho(\xi),\quad\xi= x-ct,   \label{eq37}
\end{eqnarray}
for (\ref{eq36}). The resulting ordinary differential equation is
\begin{eqnarray}
&&\left(c-F_x\right)\!\frac{\partial^2\rho}{\partial\xi^2}+(\mu-\Gamma\rho)\frac{\partial\rho}{\partial\xi} =0,  \label{eq38}
\end{eqnarray}
in which $F_x$ is the $x$-component of the chemotactic force $\mathbf{F}$. Integrating \eqref{eq38} once, we obtain
\begin{eqnarray}
\left(c-F_x\right)\!\frac{\partial\rho}{\partial\xi}+\left(\mu-\frac{\Gamma}{2}\rho\right)\!\rho =-K,  \label{eq39}
\end{eqnarray}
where $K$ is a constant. From this, we get
\begin{eqnarray}
\left(c-F_x\right)\!\frac{2}{\Gamma}\frac{\partial\rho}{\partial\xi}=\rho^2-2\frac{\mu}{\Gamma}\rho-\frac{2K}{\Gamma}.  \label{eq40}
\end{eqnarray}
Setting $\rho=\frac{\mu}{\Gamma}+\nu\tanh(\lambda\xi)$, we find $\nu^2=\frac{\mu^2+2K\Gamma}{\Gamma^2}$ and $2\nu\lambda(c-F_x)/\Gamma=-\nu^2$, thereby obtaining
\begin{eqnarray}
\rho=\frac{\mu}{\Gamma}-\frac{\sqrt{2K\Gamma+\mu^2}}{\Gamma}\tanh\!\left[\frac{\sqrt{2K\Gamma+\mu^2}}{2(c-F_x)}(\xi-\xi_0)\right]\!.  \label{eq41}
\end{eqnarray}
Here $\xi_0$ is a constant of integration. Thus $\tilde{p}=\frac{\partial\rho}{\partial t}=-c\frac{\partial\rho}{\partial\xi}$ yields
\begin{eqnarray}
\tilde{p}=\frac{(2K\Gamma+\mu^2)c}{2\Gamma(c-F_x)}\mbox{sech}^2\!\left[\frac{\sqrt{2K\Gamma+\mu^2}}{2(c-F_x)}(x-ct-\xi_0)\right]\!\!.  \label{eq42}
\end{eqnarray}
This is similar to the usual soliton solution of the Korteweg-de Vries equation except that we now have  three parameters, $c$, $K$ and $\xi_0$. Note that the soliton appears as consequence of a dominant balance of time derivative, convection, and source terms in \eqref{eq31}. The existence of the soliton solution is consequence of the quadratic anastomosis term in \eqref{eq14} first derived in \cite{bon14}. While simulations of the deterministic \cite{bon14} and stochastic descriptions \cite{ter16} clearly exhibit a soliton-like solution, the derivation presented here first appeared in \cite{bon16}.

\section{Collective coordinates}\label{sec:cc}
In this section, we shall discuss the effect of small diffusion and a slowly varying TAF concentration on the soliton. Let the soliton solution (\ref{eq42}) be written as
\begin{eqnarray}
&&\!\tilde{p}_s=\frac{(2K\Gamma+\mu^2)c}{2\Gamma(c-F_x)}\mbox{ sech}^2s,  \label{eq43}\\
&&s=\frac{\sqrt{2K\Gamma+\mu^2}}{2(c-F_x)}\,\xi, \quad\xi=x-X(t),\label{eq44}\\
&&\dot{X}=\frac{dX}{dt}=c.\label{eq45}
\end{eqnarray}
Here $X(t)$, $c(t)$ and $K(t)$ are time-dependent {\em collective coordinates} characterizing the soliton. They are supposed to vary slowly so that the marginal tip density is described by a soliton that moves and changes shape slowly according to the changes of its collective coordinates. To find equations for them, we adapt the perturbation method explained in References \cite{mer97,san14}. Note that $\tilde{p}_s$ is a function of $\xi$ and also of $\mathbf{x}$ and $t$ through $C(t,\mathbf{x})$,
\begin{equation}
\tilde{p}_s=\tilde{p}_s\!\left(\xi;K,c,\mu(C),F_x\!\left(C,\frac{\partial C}{\partial x}\right)\!\right)\!. 
\label{eq46}
\end{equation}
We assume that the time and space variations of $C$, which appear when $\tilde{p}_s$ is differentiated with respect to $t$ or $x$, produce terms that are small compared to $\partial\tilde{p}_s/\partial\xi$. As indicated in Appendix \ref{app3}, we shall consider that $\mu(C)$ is approximately constant, ignore $\partial C/\partial t$ because the TAF concentration is varying slowly (the dimensionless coefficients $\kappa$ and $\chi$ appearing in the TAF equation \eqref{eq33} are very small according to Table \ref{table2}) and ignore $\partial^2\tilde{p}_s/\partial i\partial j$, where $i,j=K,\, F_x$. Appendix \ref{app1} explains what happens if we relax these assumptions. We now insert (\ref{eq43}) and \eqref{eq44} into (\ref{eq31}), thereby obtaining 
\begin{eqnarray}
&&\left(F_x-\dot{X}\right)\!\frac{\partial\tilde{p}_s}{\partial\xi}+\frac{\partial\tilde{p}_s}{\partial K} \dot{K}+\frac{\partial\tilde{p}_s}{\partial c}\dot{c}+\tilde{p}_s\nabla_x\cdot\mathbf{F} \nonumber\\&&
+\frac{\partial\tilde{p}_s}{\partial F_x}\!\left(\frac{\partial F_x}{\partial t}+\mathbf{F}\cdot\nabla_xF_x\right)\! -\frac{1}{2\beta}\!\left(\frac{\partial^2\tilde{p}_s}{\partial\xi^2} \right.
\nonumber\\
&&\left.+2\frac{\partial^2\tilde{p}_s}{\partial\xi\partial F_x}\frac{\partial F_x}{\partial x}+ \frac{\partial\tilde{p}_s}{\partial F_x}\Delta_xF_x\right)\!=\mu\tilde{p}_s \nonumber\\&&
-\Gamma\tilde{p}_s\!\!\int_0^t\tilde{p}_s dt. \label{eq47}
\end{eqnarray}
Eq.\ (\ref{eq31}) with $1/\beta=0$ and constant $\mathbf{F}$ has the soliton solution \eqref{eq43}-\eqref{eq44}. Using this fact and \eqref{eq45}, (\ref{eq47}) becomes
\begin{eqnarray}
&&\frac{\partial\tilde{p}_s}{\partial K} \dot{K}+\frac{\partial\tilde{p}_s}{\partial c}\dot{c}=\mathcal{A}, 
\label{eq48}\\
&&\mathcal{A}=\! \frac{1}{2\beta}\frac{\partial^2\tilde{p}_s}{\partial\xi^2}\!-\!\tilde{p}_s\nabla_x\!\cdot\!\mathbf{F}\!-\!\frac{\partial\tilde{p}_s}{\partial F_x}\!\left[\mathbf{F}\!\cdot\!\nabla_xF_x\!-\!\frac{1}{2\beta}\Delta_xF_x\right]\! \nonumber\\&&\quad
+\frac{1}{\beta}\frac{\partial^2\tilde{p}_s}{\partial\xi\partial F_x}\frac{\partial F_x}{\partial x}. \label{eq49}
\end{eqnarray}
See Appendix \ref{app3} for the precise meaning of these equations.

We now find collective coordinate equations (CCEs) for $K$ and $c$. As the lump-like angiton moves on the $x$ axis, we set $y=0$ to capture the location of its maximum. On the $x$ axis, the profile of the angiton is the soliton \eqref{eq43}-\eqref{eq44}. We first multiply (\ref{eq48}) by $\partial\tilde{p}_s/\partial  K$ and integrate over $x$. We consider a fully formed soliton far from primary vessel and tumor. As it decays exponentially for $|\xi |\gg 1$, the soliton is considered to be localized on some finite interval $(-\mathcal{L}/2,\mathcal{L}/2)$. The coefficients in the soliton formulas \eqref{eq43}-\eqref{eq44} and the coefficients in \eqref{eq48} depend on the TAF concentration at $y=0$, therefore they are functions of $x$ and time and get integrated over $x$. The TAF varies slowly on the support of the soliton, and therefore we can approximate the integrals over $x$ by
\begin{eqnarray}
&&\!\int_\mathcal{I}\! F(\tilde{p}_s(\xi;x,t),x) dx \nonumber\\&&\quad\quad\quad
\approx\!\frac{1}{\mathcal{L}}\int_\mathcal{I}\!\!\left(\int_{-\mathcal{L}/2}^{\mathcal{L}/2}\! F(\tilde{p}_s(\xi;x,t),x) d\xi\!\right)\! dx. \label{eq50}
\end{eqnarray}
See Appendix \ref{app3}. The interval $\mathcal{I}$ over which we integrate should be large enough to contain most of the soliton, of extension $\mathcal{L}$. Thus the CCEs hold only after the initial soliton formation stage. Near the tumor, the boundary condition affects the soliton and we should exclude an interval near $x=1$ from $\mathcal{I}$. We shall specify the integration interval $\mathcal{I}$ in the next section. Acting similarly, we multiply \eqref{eq48} by $\partial\tilde{p}_{s}/\partial c$ and integrate over $x$. From the two resulting formulas, we then find $\dot{K}$ and $\dot{c}$ as fractions. The factors $1/\mathcal{L}$ cancel out from their numerators and denominators. As the soliton tails decay exponentially to zero, we can set $\mathcal{L}\to\infty$ and obtain the following CCEs \cite{bon16}
\begin{widetext}
\begin{eqnarray}
&&\dot{K}=\frac{\int_{-\infty}^\infty \frac{\partial\tilde{p}_s}{\partial K}\mathcal{A}d\xi\int_{-\infty}^\infty\!\!\left(\frac{\partial\tilde{p}_s}{\partial c}\right)^2\!\!d\xi\!- \int_{-\infty}^\infty \frac{\partial\tilde{p}_s}{\partial c}\mathcal{A}d\xi\int_{-\infty}^\infty\!\frac{\partial\tilde{p}_s}{\partial K}\frac{\partial\tilde{p}_s}{\partial c} d\xi\!}{\int_{-\infty}^\infty\!\!\left(\frac{\partial\tilde{p}_s}{\partial K}\right)^2\!\!d\xi\!\int_{-\infty}^\infty\!\!\left(\frac{\partial\tilde{p}_s}{\partial c}\right)^2\!\!d\xi\!-\left(\int_{-\infty}^\infty \frac{\partial\tilde{p}_s}{\partial c}\frac{\partial\tilde{p}_s}{\partial K}d\xi\right)^2},\label{eq51}\end{eqnarray}
\begin{eqnarray}
&&\dot{c}=\frac{\int_{-\infty}^\infty \frac{\partial\tilde{p}_s}{\partial c}\mathcal{A}d\xi\int_{-\infty}^\infty\!\!\left(\frac{\partial\tilde{p}_s}{\partial K}\right)^2\!\!d\xi\!- \int_{-\infty}^\infty \frac{\partial\tilde{p}_s}{\partial K}\mathcal{A}d\xi\int_{-\infty}^\infty\!\frac{\partial\tilde{p}_s}{\partial K}\frac{\partial\tilde{p}_s}{\partial c} d\xi\!}{\int_{-\infty}^\infty\!\!\left(\frac{\partial\tilde{p}_s}{\partial K}\right)^2\!\!d\xi\!\int_{-\infty}^\infty\!\!\left(\frac{\partial\tilde{p}_s}{\partial c}\right)^2\!\!d\xi\!-\left(\int_{-\infty}^\infty \frac{\partial\tilde{p}_s}{\partial c}\frac{\partial\tilde{p}_s}{\partial K}d\xi\right)^2}.  \label{eq52}\end{eqnarray}
\end{widetext}
In these equations, all terms varying slowly in space have been averaged over the interval $\mathcal{I}$. The last term in \eqref{eq49} is odd in $\xi$ and does not contribute to the integrals in \eqref{eq51} and \eqref{eq52} whereas all other terms in \eqref{eq49} are even in $\xi$ and do contribute. The integrals appearing in (\ref{eq51}) and (\ref{eq52}) are calculated in Appendix \ref{app4}. The resulting CCEs are
\begin{eqnarray}
&&\dot{K}= \frac{(2K\Gamma\!+\!\mu^2)^2}{4\Gamma\beta(c\!-\!F_x)^2}\frac{\frac{4\pi^2}{75}\!+\!\frac{1}{5}\!+\!\!\left(\frac{2F_x}{5 c}\!-\!\frac{2\pi^2}{75}\!-\!\frac{9}{10}\right)\!\!\frac{F_x}{c}}{\left(1-\frac{4\pi^2}{15}\right)\!\left(1-\frac{F_x}{2c}\right)^2}  \nonumber\\&&\quad
-\frac{2K\Gamma+\mu^2}{2\Gamma c\!\left(1-\frac{F_x}{2c}\right)}\!\left(c\nabla_x\!\cdot\mathbf{F}\! +\!\mathbf{F}\!\cdot\!\nabla_x F_x-\frac{\Delta_xF_x}{2\beta}\right)\!, \label{eq53}
\end{eqnarray}
\begin{eqnarray}
\dot{c}=-\frac{7(2K\Gamma+\mu^2)}{20\beta(c-F_x)}\frac{1-\frac{4\pi^2}{105}}{\left(1-\frac{4\pi^2}{15}\right)\!\left(1-\frac{F_x}{2c}\right)\!}  \nonumber\\
+\frac{\mathbf{F}\!\cdot\!\nabla_x F_x -(c-F_x)\nabla_x\!\cdot\!\mathbf{F}-\frac{\Delta_xF_x}{2\beta}}{2-\frac{F_x}{c}},
\label{eq54}
\end{eqnarray}
in which the functions of $C(t,x,y)$ have been averaged over the interval $\mathcal{I}$ and we have set $y=0$. We expect the CCEs \eqref{eq53}-\eqref{eq54} to describe the mean behavior of the soliton whenever it is far from primary vessel and tumor. We back this point of view by the numerical simulations reported in the next section.

\section{Numerical results}
\label{sec:numerical}
Based on numerical simulations \cite{bon16}, we expect that the vessel tip density approaches the soliton after some time. Initially there are few tips and the density is small so that the nonlinear anastomosis terms in \eqref{eq14} or in \eqref{eq31} are small. Tips proliferate and the anastomosis terms kick in. The soliton formation should be described as the solution of a semi-infinite initial-boundary value problem. Ideally, we would match the solution of the soliton formation stage with a stage of a soliton moving far from boundaries, which is the crucial stage described by Equations \eqref{eq53}-\eqref{eq54} for the collective coordinates. We expect the soliton solution to be an asymptotically stable solution of the vessel tip density equation \eqref{eq14} on the whole 1D real line and also for the 2D slab geometry considered in this paper (provided the primary vessel is at $x=-\infty$ and the tumor is at $x=+\infty$). For a slowly varying TAF density, the stable soliton will instantaneously adapt its shape and velocity according to the solution of the CCEs \eqref{eq53}-\eqref{eq54}. 

In this paper, we will solve numerically the full equations \eqref{eq14} (with $q=1$) and \eqref{eq15} for the vessel tip density and the TAF density (deterministic description), which we will also obtain by ensemble averages from stochastic simulations as explained in \cite{ter16}. From these simulations, we will obtain the evolution of the soliton collective coordinates thereby reconstructing the marginal tip density at $y=0$ from \eqref{eq43}. The soliton provides a simple description of tumor induced angiogenesis that agrees with numerical simulations of the stochastic process and with numerical simulations of the deterministic description. 

Both deterministic or stochastic simulations show that the soliton is formed after some time $t_0=0.2$ (10 hours) following angiogenesis initiation. To find the soliton evolution afterwards, we need to solve the CCEs \eqref{eq53}-\eqref{eq54} whose coefficients are spatial averages over a certain interval $x\in\mathcal{I}$ that depend on the TAF concentration $C(t,x,y)$ and its derivatives calculated at $y=0$. The interval $\mathcal{I}$ should exclude regions affected by boundaries. We calculate the spatially averaged coefficients in \eqref{eq53}-\eqref{eq54} by: (i) approximating all differentials by second order finite differences, (ii) setting $y=0$, and (iii) averaging the coefficients from $x=0$ to 0.6 by taking the arithmetic mean of their values at all grid points in the interval $\mathcal{I}=(0,0.6]$. For $x>0.6$, the boundary condition at $x=1$ influences the outcome and therefore we leave values for $x>0.6$ out of the averaging. 

\begin{figure}[h]
\begin{center}
\includegraphics[width=6.0cm]{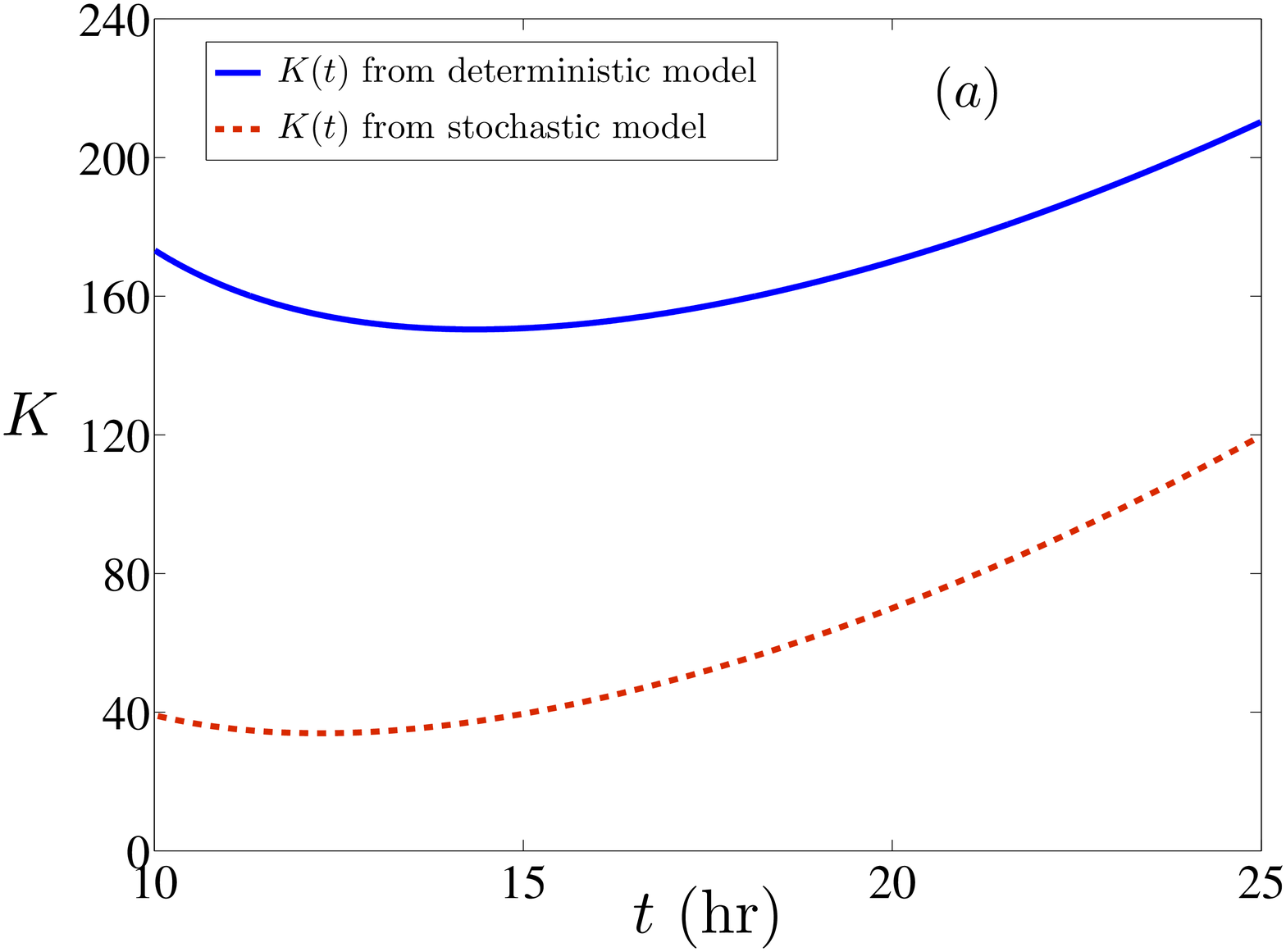} 
\includegraphics[width=6.0cm]{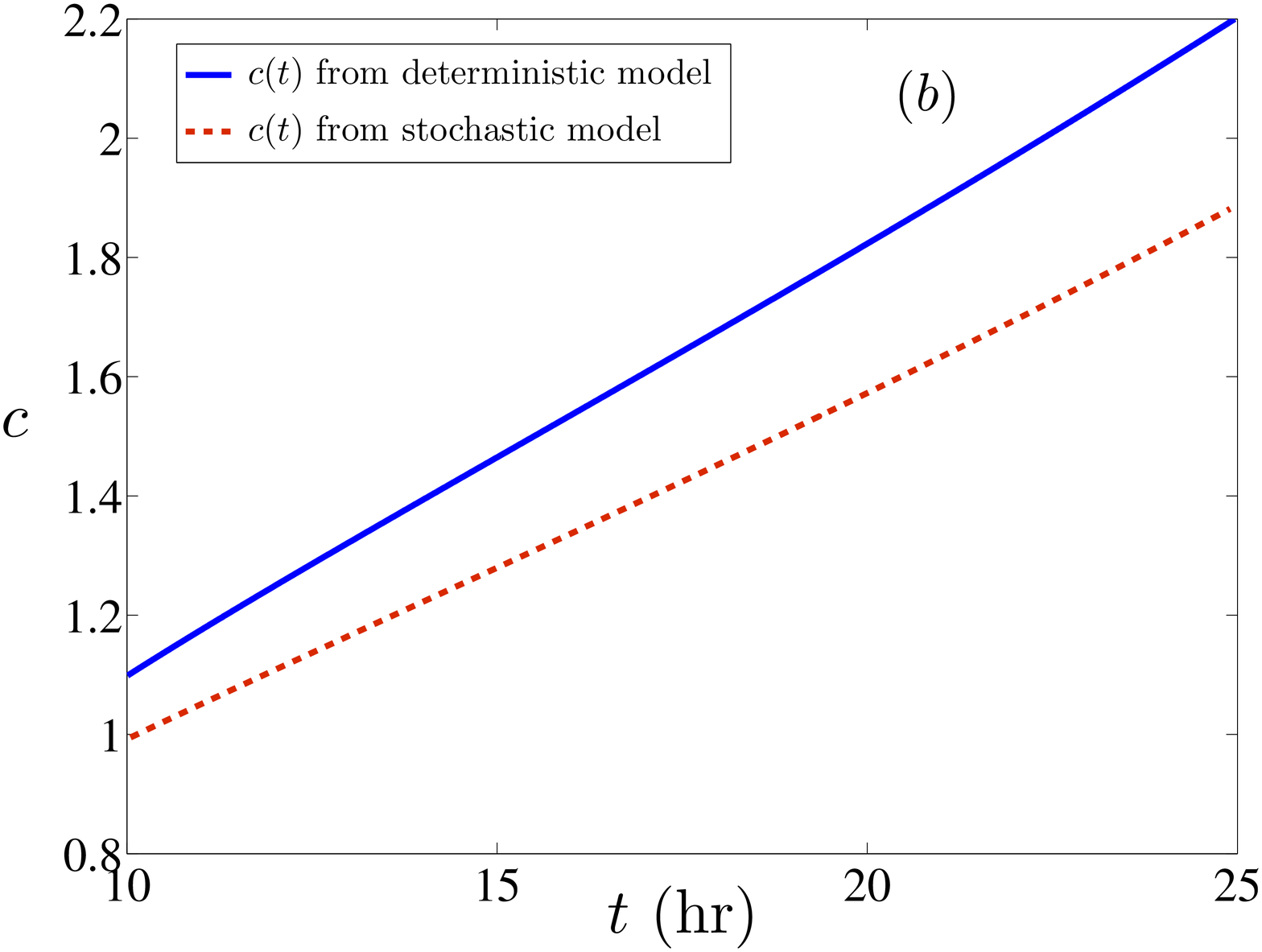} 
\includegraphics[width=6.0cm]{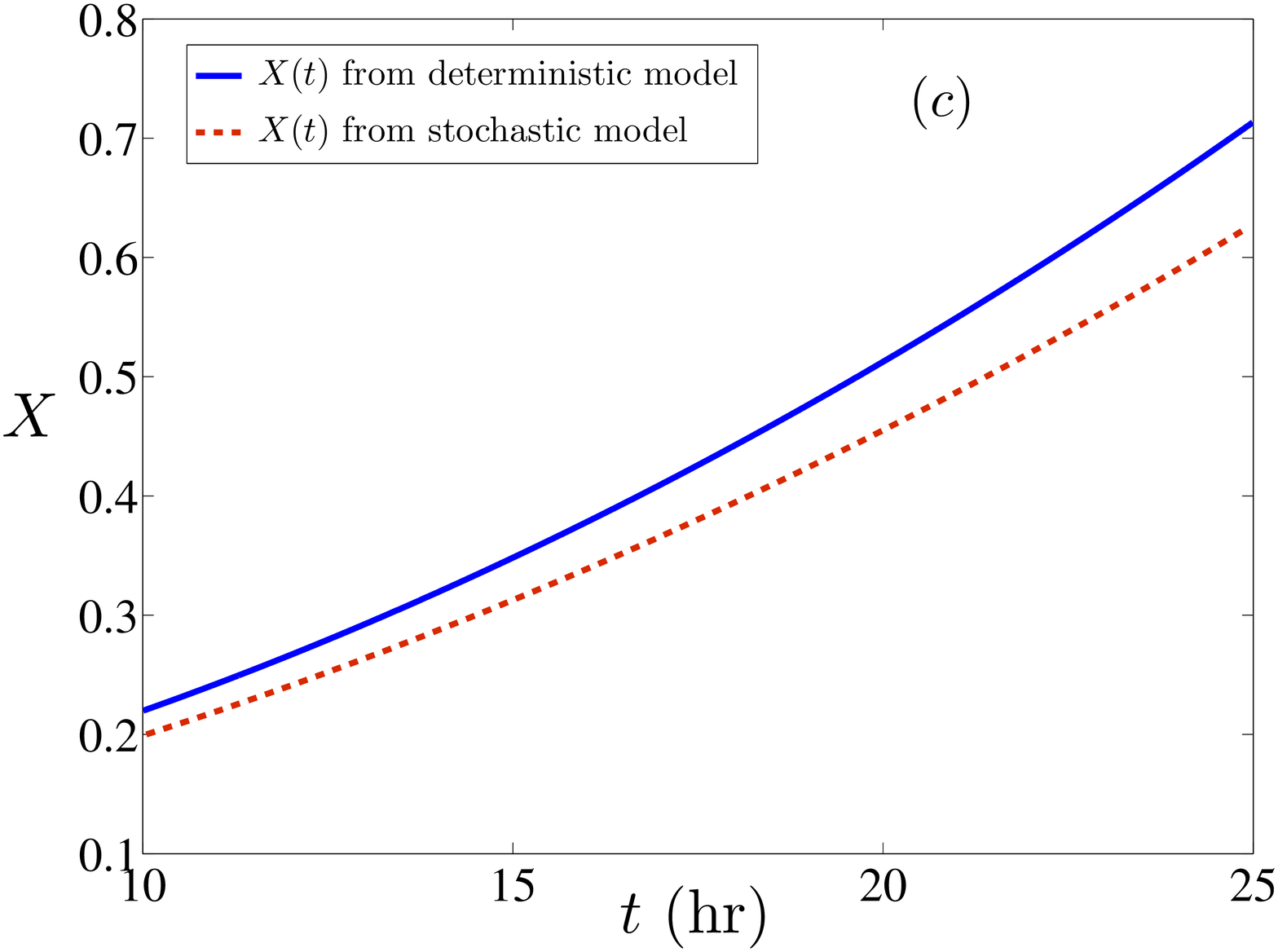}
\end{center}
\vskip-3mm 
\caption{ Evolution of the collective coordinates: (a) $K(t)$, (b) $c(t)$, and (c) $X(t)$. \label{fig1}}
\end{figure}

The initial conditions for the CCEs \eqref{eq45}, \eqref{eq53} and \eqref{eq54} are set as follows. $X(t_0)=X_0$ is the  location of the marginal tip density maximum, $\tilde{p}(t_0,x=X_0,0)$. We find $X_0=0.22$ from the deterministic description and $X_0=0.2$ from the stochastic description. We set $c(t_0)=c_0=X_0/t_0$. $K(t_0)=K_0$ is determined so that the maximum marginal tip density at $t=t_0$ coincides with the soliton peak. This yields $K_0=173$ (deterministic description) and 39 (stochastic description). Solving the CCEs \eqref{eq45}, \eqref{eq53} and \eqref{eq54} with these initial conditions, we obtain the curves depicted in Figure \ref{fig1}.

\begin{figure}[h]
\begin{center}
\includegraphics[width=9.0cm]{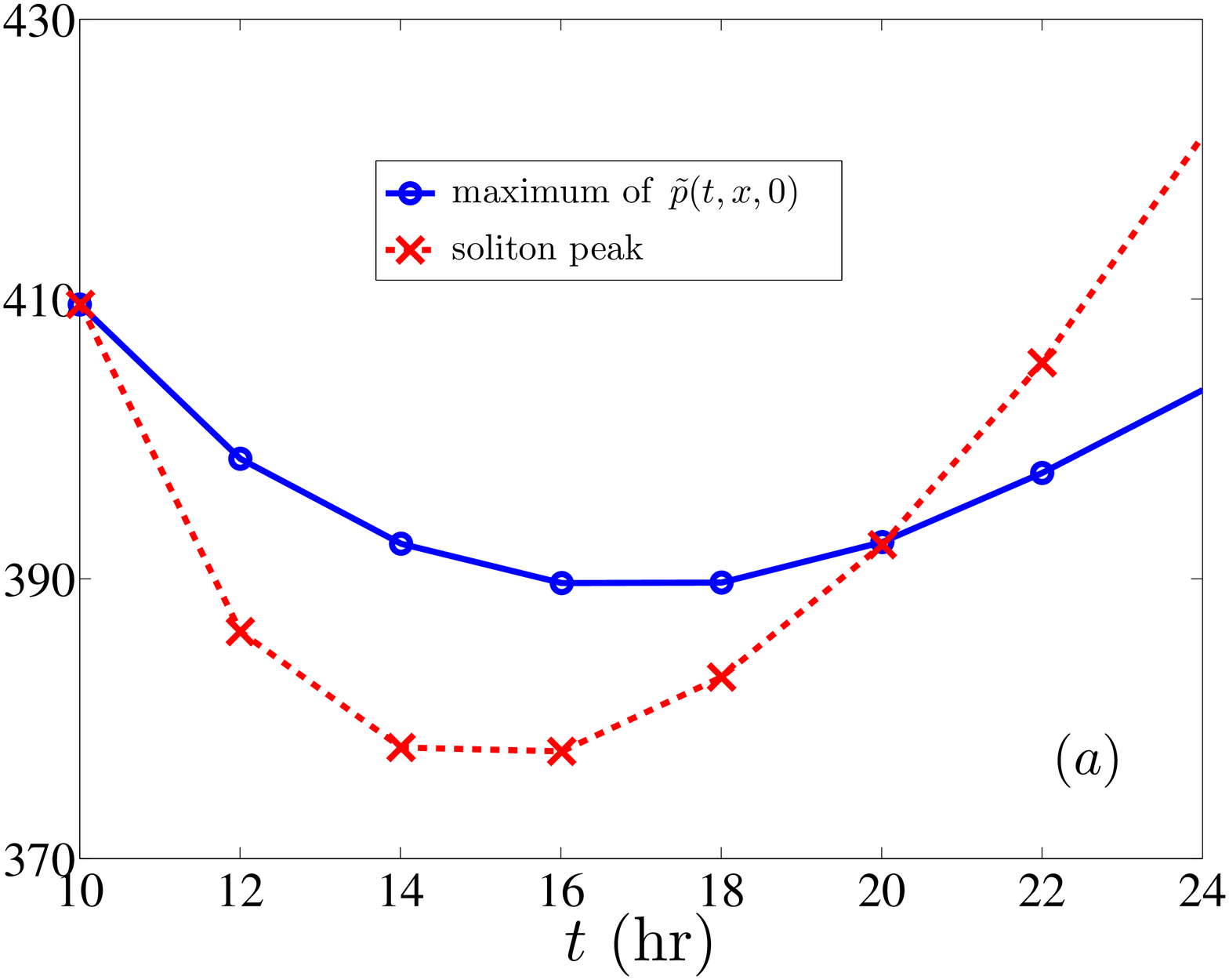} 
\includegraphics[width=9.0cm]{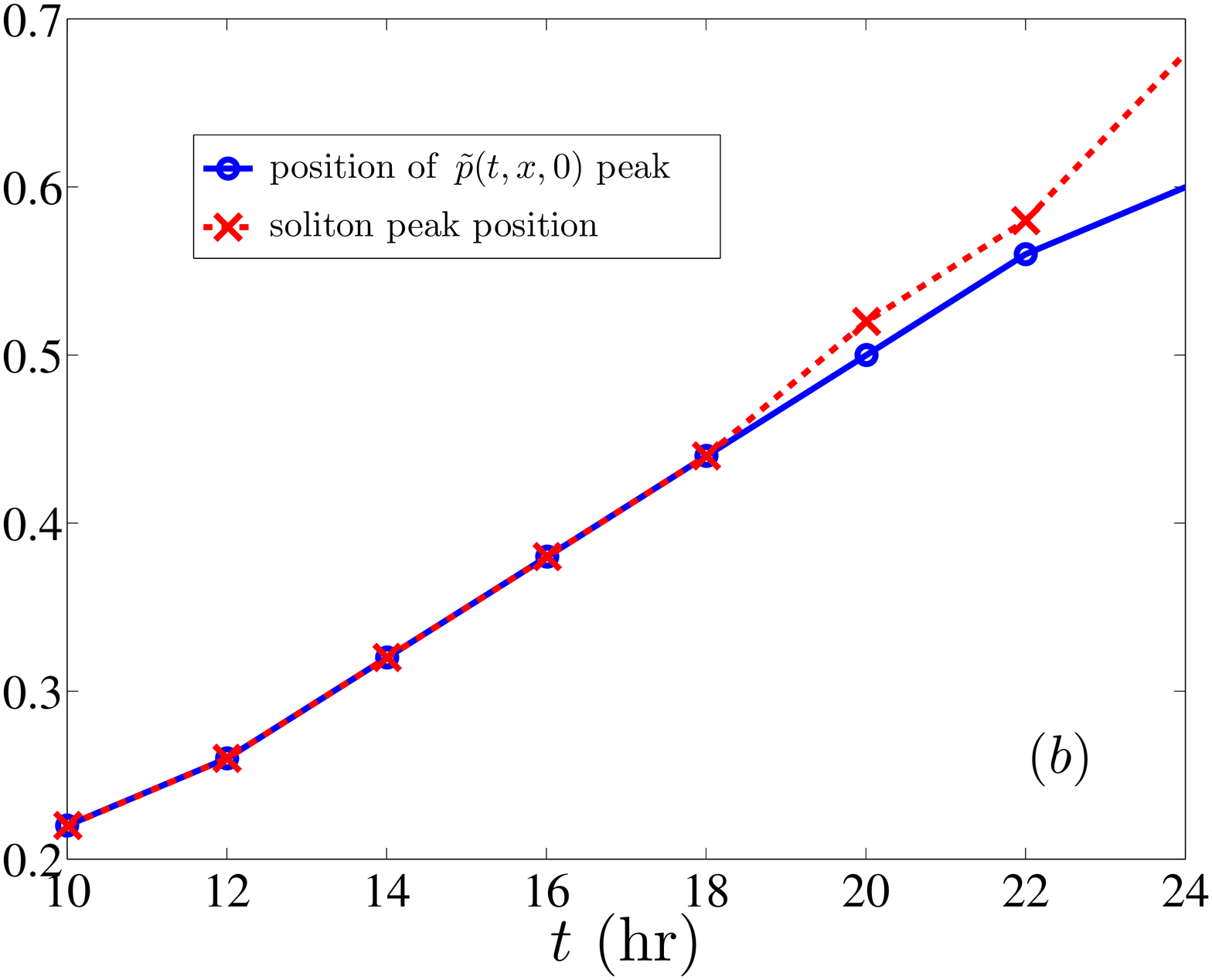}
\end{center}
\vskip-4mm \caption{Deterministic description: Comparison between the maximum value of  $\tilde{p}(t,x,0)$ and its value as predicted by soliton collective coordinates. (a) Evolution of the maximum value of the marginal tip density (relative error smaller than 4.5\%). (b) Evolution of the position of the maximum marginal tip density on $[0,1]$ (at $t=20$ and 22 h, the absolute error is the space step in the numerical method, $\Delta x = 0.02$; at $t=24$ h, the error is $4\Delta x$). \label{fig2}}
\end{figure}
\begin{figure}[h]
\begin{center}
\includegraphics[width=9.0cm]{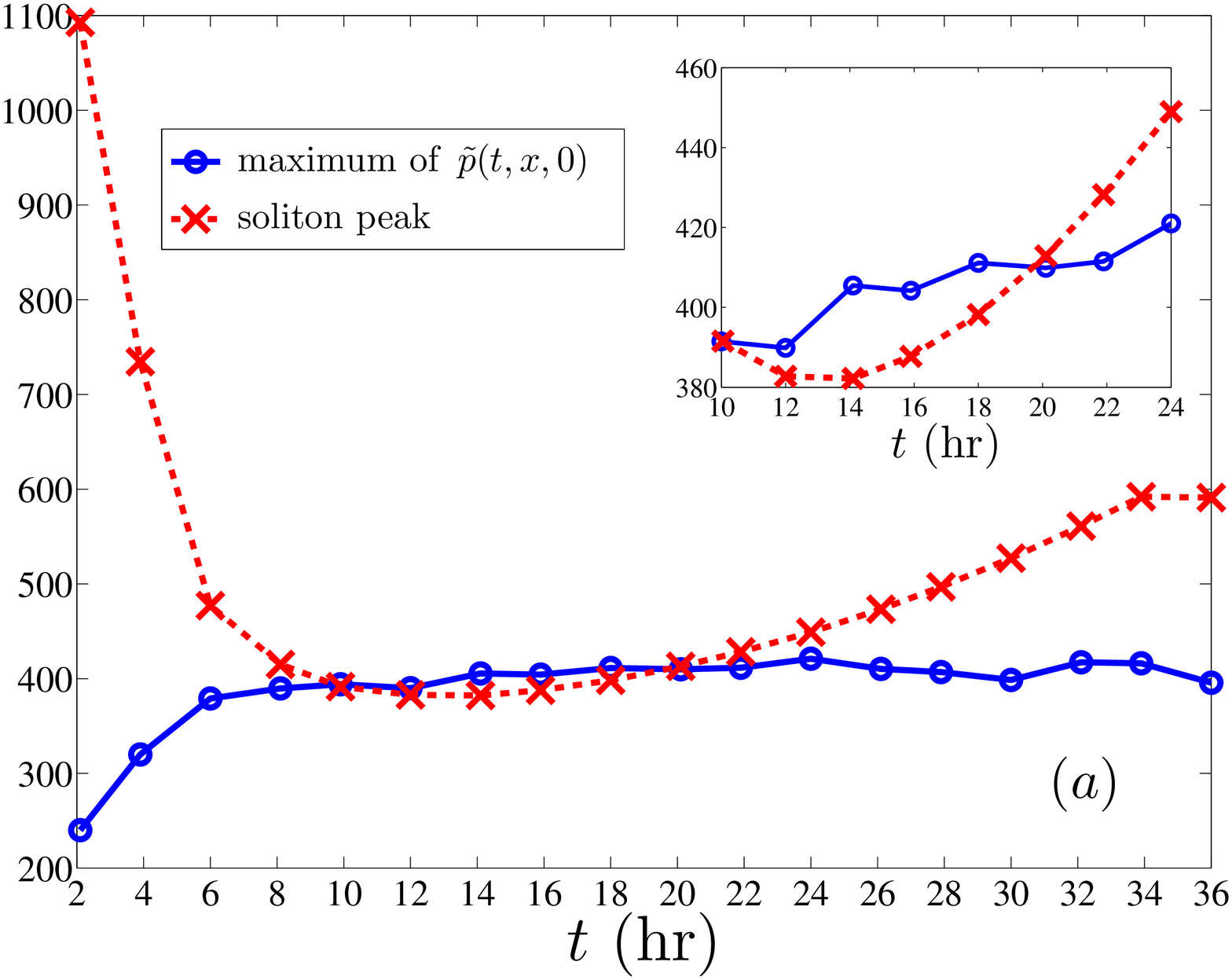} 
\includegraphics[width=9.0cm]{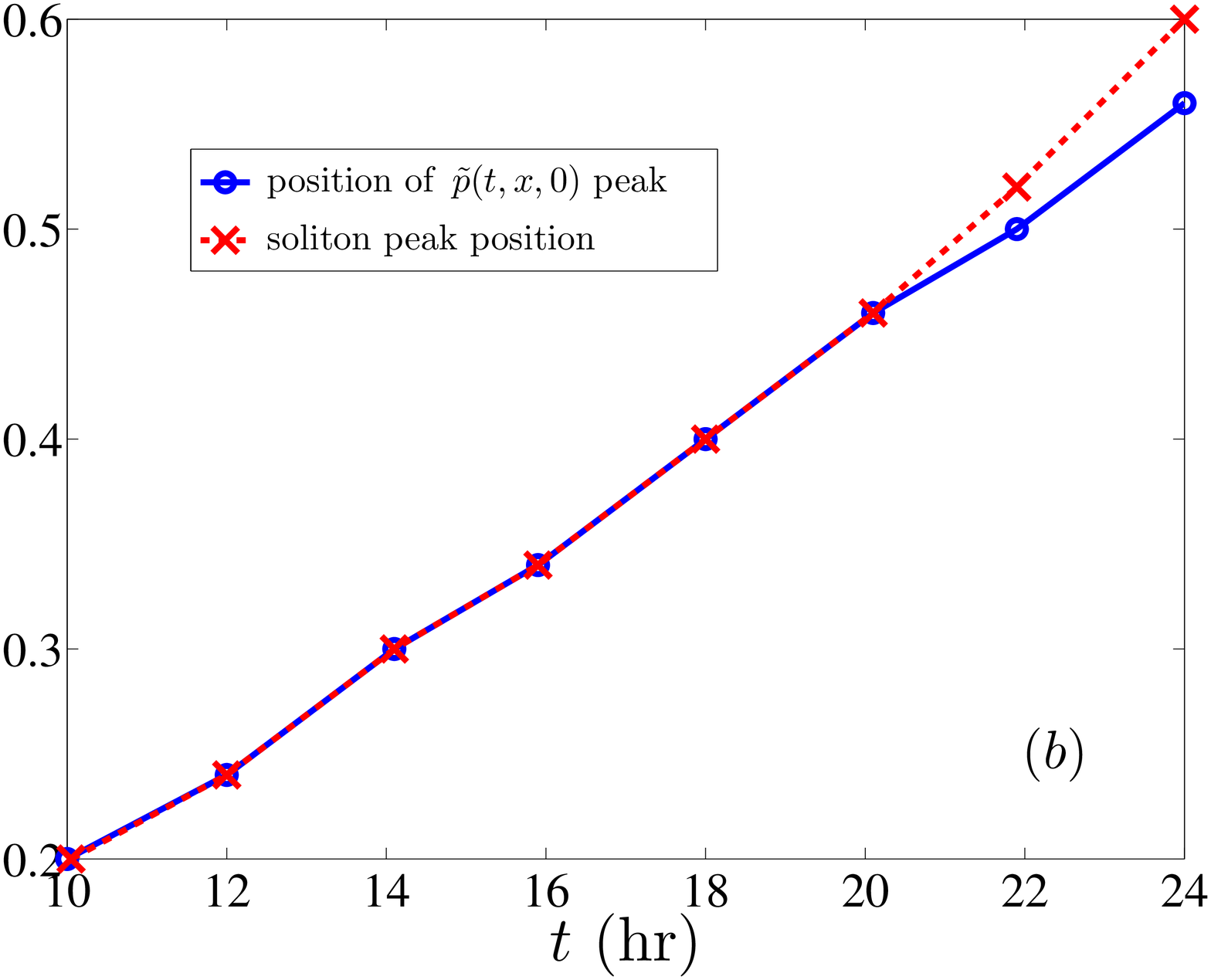}
\end{center}
\vskip-4mm \caption{Same as in Figure \ref{fig2} for the stochastic description. The zoom in Figure \ref{fig3}(a) corresponds to Figure \ref{fig2}(a) but we have drawn the same figure with a larger time span to show more clearly the time interval over which the soliton approximates the maximum marginal tip density. The relative error is smaller than $6.7\%$ for the maximum marginal tip density (calculated by ensemble average over 400 realizations \cite{ter16}), whereas the error in the predicted position of the maximum marginal tip density is $\Delta x = 0.02$ at 22h and $2\Delta x$ at 24 h. \label{fig3}}
\end{figure}

\begin{figure}[htbp]
\begin{center}
\end{center} 
\includegraphics[width=10cm]{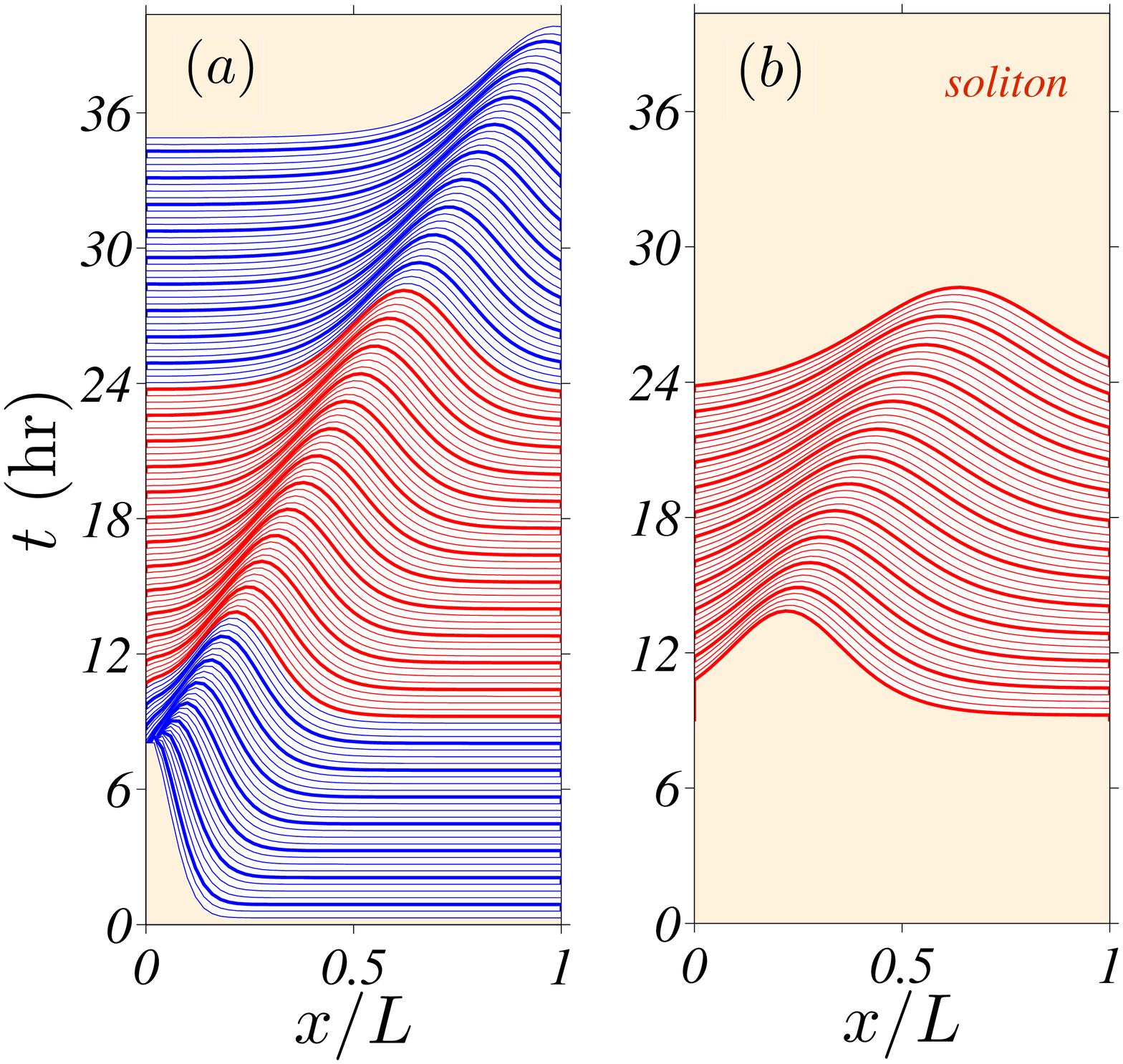}\\ 
\includegraphics[width=10cm]{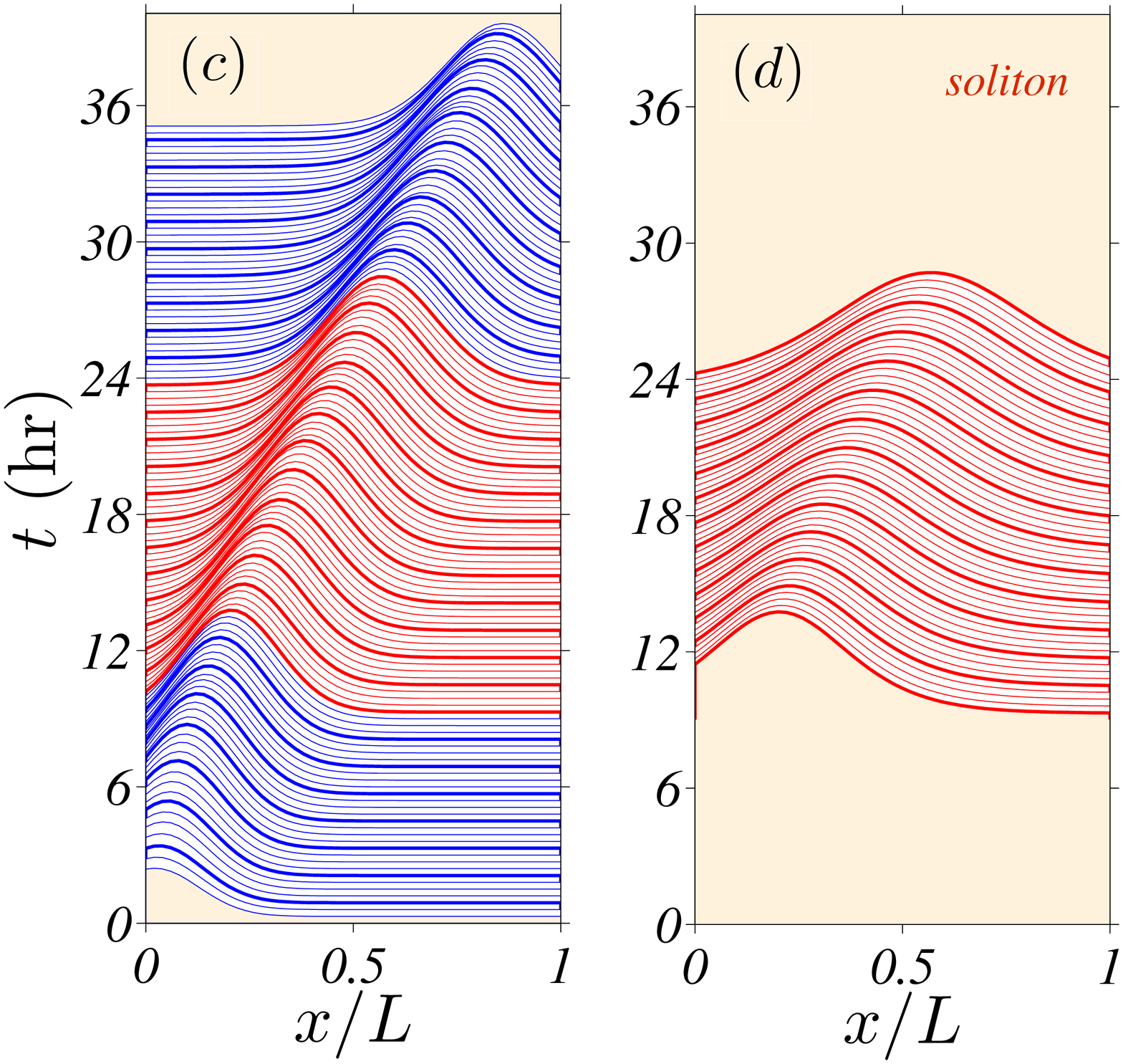}
\caption{Comparison of the marginal tip density profile to that of the moving soliton for (a) and (b): Deterministic description; (c) and (d): Stochastic description averaged over 400 replicas.
\label{fig4}}
\end{figure}

Using the soliton collective coordinates depicted in Figure \ref{fig1} and \eqref{eq43}-\eqref{eq44}, we reconstruct the marginal vessel tip density and find its maximum value and the location thereof for all times $t>t_0$. Figure \ref{fig2} shows that the soliton as predicted from the CCEs \eqref{eq45}, \eqref{eq53} and \eqref{eq54} compares very well with the tip density obtained by direct numerical simulation of the deterministic equations. An alternative way to find the coefficients of the CCEs and their proper initial conditions is to use ensemble averages of the stochastic process. Figure \ref{fig3} shows that such reconstruction of the soliton agrees very well with the vessel tip density provided by ensemble averages of the stochastic process during the 14 hour time interval when soliton motion is not affected by boundaries. There is a large discrepancy between the maximum marginal tip density as predicted by the soliton and by the stochastic process during the first 10 hours of angiogenesis, which clearly marks the duration of the initial stage of soliton formation. After this stage, we note that the location of the maximum of the marginal tip density is very closely predicted by the location of the soliton peak as a function of time, both by using ensemble averages of the stochastic process as in Figure \ref{fig3} or by solving numerically the deterministic description as in Figure \ref{fig2}. This is also clearly shown in the reconstruction of the soliton marginal tip density depicted in Figure \ref{fig4}.

\begin{figure}[h]
\begin{center}
\begin{tabular}{c}
\includegraphics[width=9.0cm]{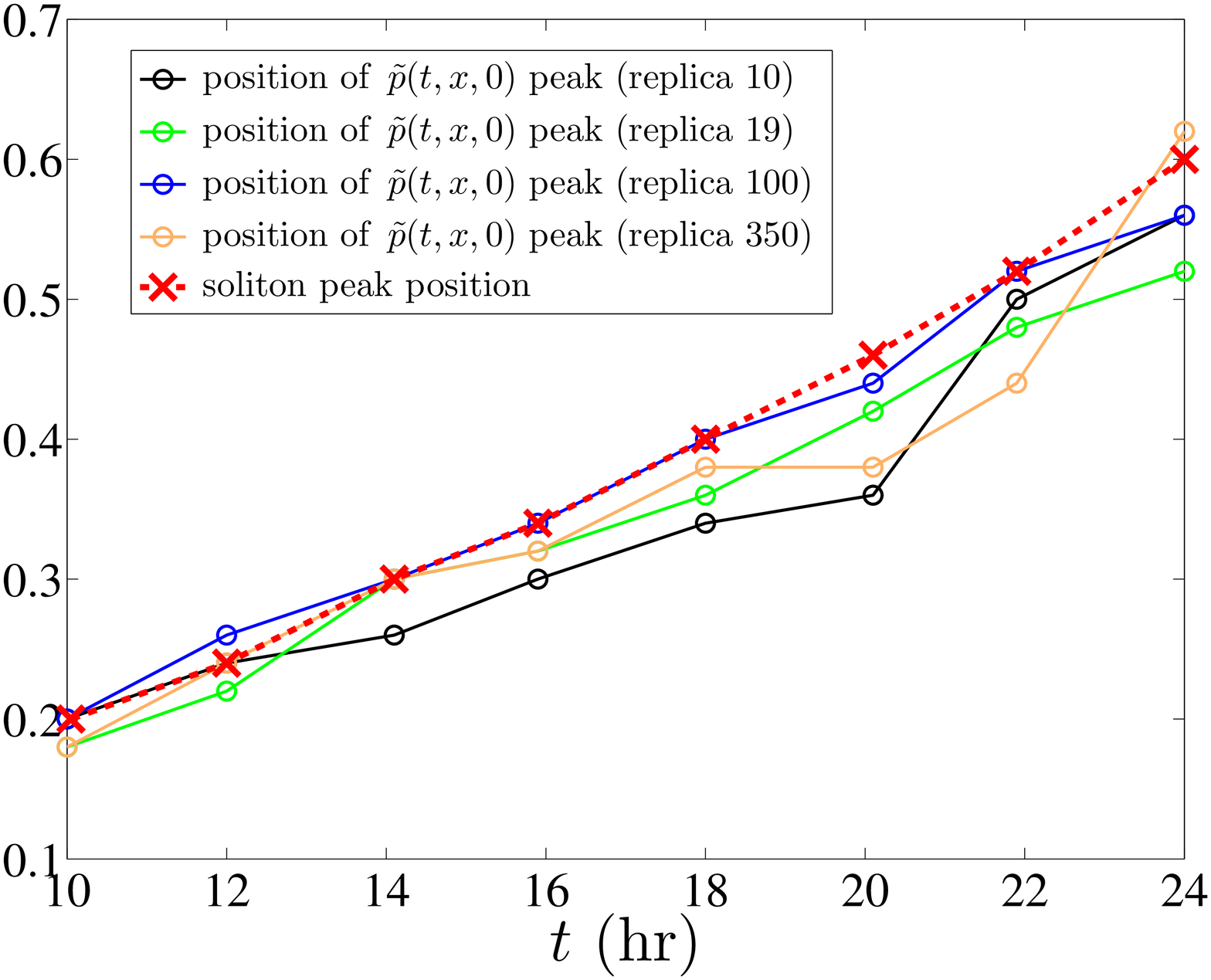}
\end{tabular}
\end{center}
\vskip-4mm \caption{ Position of the soliton peak density compared to that of the maximum marginal tip density for different replicas of the stochastic process. \label{fig5}}
\end{figure}

So far, our reconstructions have been based on ensemble averages or, what is quite similar, the marginal tip density as given by the deterministic description. In past work \cite{ter16}, we have shown that fluctuations about the mean are large and therefore the stochastic process is not self-averaging for a single realization: anastomosis precludes the formation of a large number of active tips that may enforce mean-field behavior. However a deterministic description is still possible for averages over a sufficiently large number of realizations of the stochastic process (four hundred realizations suffice), as explained extensively in \cite{ter16}. This raises an important question: {\em How well do these ensemble averages and the soliton construction represent single replicas of the stochastic process?} Figure \ref{fig5} gives a positive answer for the location of the soliton peak: The position of the soliton peak is a good approximation to the location of the maximum marginal tip density for different replicas of the stochastic process. While vessel networks may differ widely from replica to replica, the position of the maximum marginal tip density is about the same for different replicas. As the maximum of the marginal tip density is a good measure of the advancing vessel network, the soliton peak location also characterizes it. The existence of other seemingly self-averaging quantities related to the soliton is an open question.

\begin{figure}[h]
\begin{center}
\begin{tabular}{c}\hskip -0.4cm
\includegraphics[width=9.0cm]{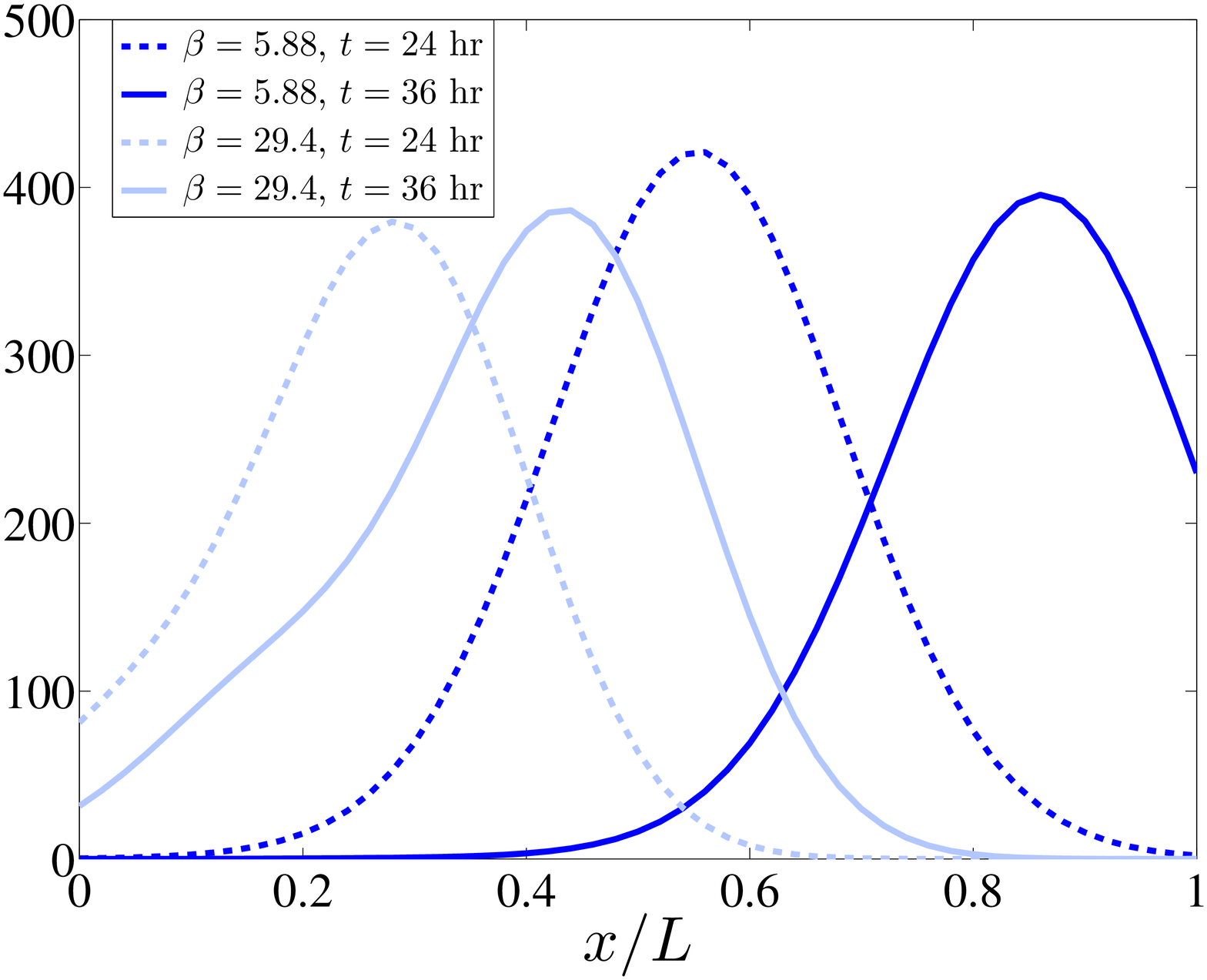}
\end{tabular}
\end{center}
\vskip -0.6cm \caption{ Marginal vessel tip density profiles at 24 (dashed lines) and 36 hours (solid lines) for $\beta=5.88$ (blue lines), and $\beta=29.4$ (red lines).  \label{fig6}}
\end{figure}

\begin{figure}[h]
\begin{center}
\begin{tabular}{c}\hskip -0.4cm
\includegraphics[width=9.0cm]{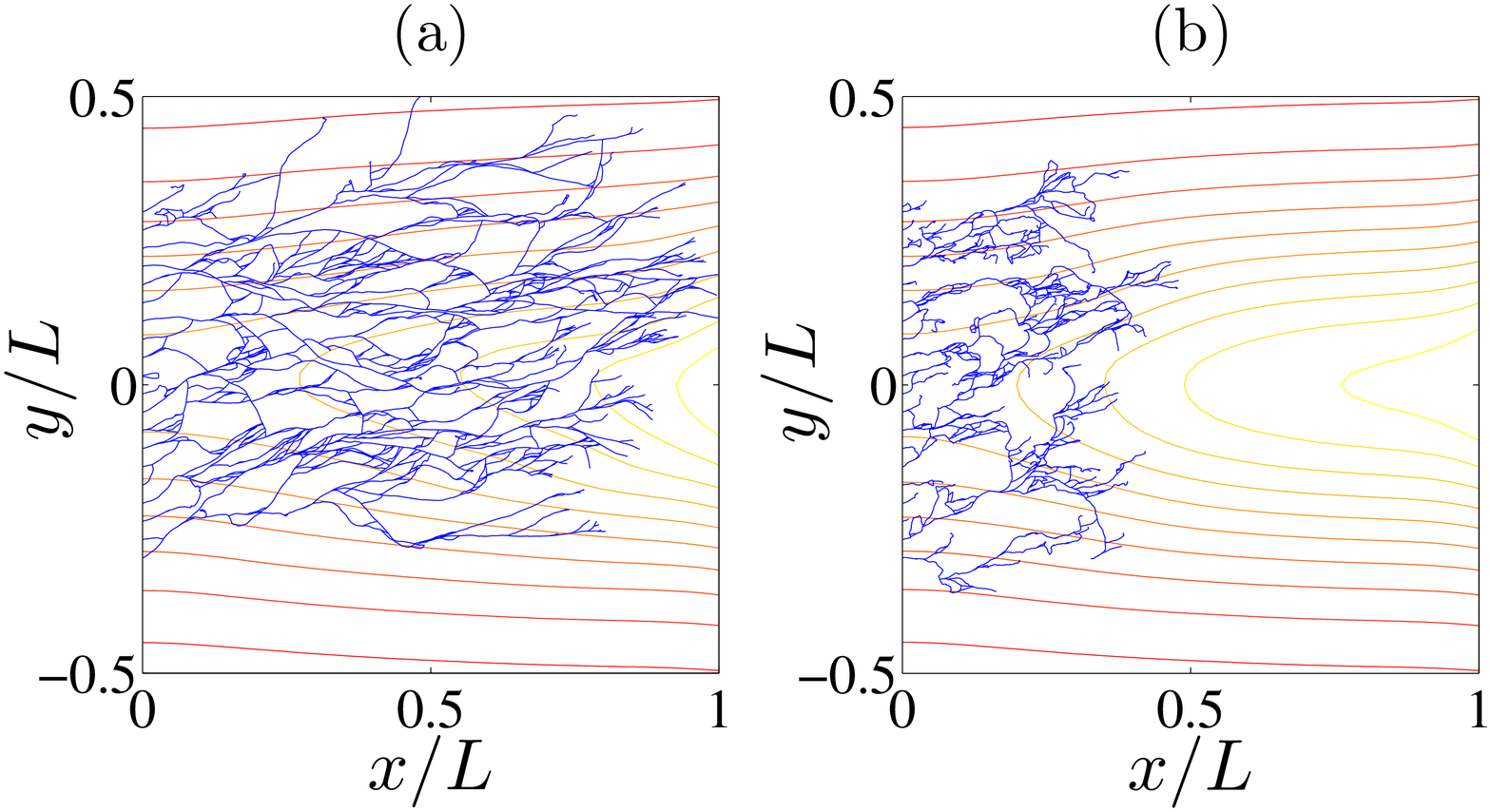}
\end{tabular}
\end{center}
\vskip -1.6cm \caption{Comparison between the vessel networks of two replicas after 36 hours for (a) $\beta=5.88$, and (b) $\beta=29.4$. The TAF level curves have also been depicted. \label{fig7}}
\end{figure}

We can use the soliton construction as a simple means to evaluate the influence of new mechanisms on angiogenesis. For instance, suppose that some drug causes the friction coefficient $\beta$ to increase fivefold. Then the marginal tip density gets delayed as shown by Figure \ref{fig6}. This can be evaluated easily and cheaply by solving the CCEs. What does this mean for replicas of the angiogenesis process? Figure \ref{fig7} displays the vessel networks formed after 36 hours for $\beta = 5.88$ and 29.4 in two different replicas of the stochastic process. For $\beta = 5.88$, the vessel network of one replica of the angiogenesis process has reached the tumor at $x=1$ after 36 hours, for $\beta= 29.4$ the vessel network is only half way through its road to the tumor after that time. Had the increase in $\beta$ been the result of some therapy, we could have ascertained its merits by solving the CCEs and inferring the arrest of the vessel network from the result. 

\section{Conclusions}
\label{sec:conclusions}
Previous work has shown that a simple stochastic model of tumor induced angiogenesis could be described deterministically by an integrodifferential equation of Fokker-Planck type with a linear birth term and a nonlinear death (anastomosis) term \cite{bon14,ter16}. Anastomosis keeps the number of vessel tips rather small (about one hundred) and therefore the vessel tip density has to be reconstructed from ensemble averages of the stochastic process, which is not self-averaging. Numerical simulations of stochastic and deterministic equations show that the vessel tip density advances from the primary vessel towards the tumor as a stable moving lump or angiton whose profile along the $x$ axis is soliton-like \cite{bon14,ter16,bon16}. An analytic formula for the longitudinal profile of the angiton (called the ``soliton'' in this paper) can be deduced by ignoring spatio-temporal variation of the tumor angiogenic factor and diffusion \cite{bon16}. This formula involves two collective coordinates that characterize the shape and velocity of the soliton \cite{bon16}. 

In the present work, we have derived the reduced equation for the marginal tip density by means of a Chapman-Enskog method. We have deduced the differential equations for the collective coordinates whose terms involve spatial averages over the fully grown soliton far from the tumor. We can deduce these equations both from the deterministic description and from ensemble averages of the full stochastic model. In both cases, the soliton provides a good reconstruction of the deterministic marginal tip density or its version based on ensemble averages, provided the soliton is not too close to the tumor. As said before, fluctuations are large because anastomosis keeps a small number of active vessel tips at all time. Nevertheless, we have shown that the position of the maximum marginal tip density as given by the soliton is quite close to that given by any replica of the stochastic angiogenesis process. This indicates that the simple soliton construction yields good predictions of the evolution of the blood vessel network. 

There are mechanisms not included in our stochastic conceptual model of angiogenesis. However, many mechanisms such as haptotaxis can be included by adding terms to the force $\mathbf{F}$ in the Langevin equation for the vessel tips that depend on additional continuum fields (fibronectin, matrix degrading enzymes, etc; see e.g., \cite{cap09}). The effects of anti-angiogenic factors could be treated by including additional reaction-diffusion equations and their effects on the vessel tips \cite{lev01}. Such terms can be straightforwardly incorporated to the equations for the soliton collective coordinates using the same methodology as explained in the present paper. There are other models that postulate reinforced random walks \cite{and98,pla03,man04} or cellular Potts models with Monte Carlo dynamics \cite{bau07,sci11,hec15} instead of Langevin equations to describe the extension of vessel tips. Insofar as Fokker-Planck equations can be derived from master equations in appropriate limits \cite{gar10} and branching and anastomosis are similar to those of our conceptual model, we could use the same methodology as in the present paper to study such models. Let us recall that the soliton solution comes through a balance of birth and death terms, convection and time derivative terms in the equation for the marginal tip density. These terms would also appear in special limits of the random walk or cellular Potts models. We consider the work presented in this paper a blueprint for using the soliton methodology to analyze more complex angiogenesis models and a first step to control angiogenesis through soliton dynamics. 

\acknowledgments
We thank Vincenzo Capasso, Bjorn Birnir and Boris Malomed for fruitful discussions. This work has been supported by the Ministerio de Econom\'\i a y Competitividad grant  MTM2014-56948-C2-2-P.   

\appendix
\setcounter{equation}{0}
\renewcommand{\theequation}{A.\arabic{equation}}
\section{Boundary conditions for the deterministic equations}\label{app2}
The nondimensional boundary conditions for the TAF are \cite{bon14}
\begin{eqnarray}\label{b1}
\frac{\partial C}{\partial x}(t,0,y)=0, \,\,\frac{\partial C}{\partial x}(t,1,y)=\frac{aL}{d_2C_R} e^{-y^2L^2/b^2}\!
\end{eqnarray}
($b$ is half the tumor width) and $\lim_{y\to\pm\infty}C= 0$. We do not intend to follow the process of angiogenesis beyond the time that vessels tip have arrived at the tumor and therefore we do not give the latter a finite length. We use a Gaussian as the initial condition for the TAF 
\begin{eqnarray}
C(0,x,y)=1.1\, e^{-[(x-1)^2L^2/c^2+y^2L^2/b^2]},\label{b2}
\end{eqnarray} 
for appropriate $b$ and $c$. The boundary conditions for the tip density are \cite{bon14}
\begin{eqnarray}
&&\!\! p^+(t,0,y,v,w)\!
=\!\frac{e^{-|\mathbf{v}-\mathbf{v}_0|^2}}{\int_0^{\infty}\!\int_{-\infty}^{\infty}
 v' e^{-|\mathbf{v}'-\mathbf{v}_0|^2}dv'\,dw'} \nonumber \\
&&\!\! \times\! \!\left[j_0(t,y)\! -\!\! \int_{-\infty}^0\!\int_{-\infty}^{\infty}\!\!
v' p^-(t,0,y,v',w')d v' dw'\!\right]\!\!, \label{b3} \\
&&\!\!p^-(t,1,y,v,w)\!=\!\frac{e^{-|\mathbf{v}-\mathbf{v}_0|^2}}{\int_{-\infty}^0\!\int_{-\infty}^{\infty}
e^{-|\mathbf{v}'-\mathbf{v}_0|^2}dv'\,dw'} \nonumber\\
&& \!\!\times\!  \!\left[\tilde{p}(t,1,y)\! -\!\!\int_0^{\infty}\!\!\int_{-\infty}^{\infty}\! p^+(t,1,y,v',w')dv' dw'\!\right]\!\!, \label{b4}\\
&&\!\! p(t,\mathbf{x},\mathbf{v})\to 0 \mbox{ as } |\mathbf{v}|\to \infty,\label{b5}
\end{eqnarray}
where $p^+=p$ for $v>0$ and $p^-=p$ for $v<0$, $\mathbf{v}=(v,w)$. An absorbing boundary condition $p=0$ on the tumor surface would be more realistic than \eqref{b4}. However this would be computationally more costly as we would need to include a slab that extends beyond $x=1$. However the difference with the present results would be appreciable at the last stage when the vessel tips arrive at the tumor, something we do not study specifically in the present paper. In \eqref{b3}, the tip flux density at $x=0$ is \cite{bon14}
\begin{equation}
j_0(t,y)=v_0\alpha(C(t,0,y))\, p(t,0,y,v_0,w_0)\,\theta(\tau-t), \label{b6}
\end{equation}
for the vector velocity $\mathbf{v}_0=(v_0,w_0)$, with $|\mathbf{v}_0|=1$. Different from \cite{bon14}, we have included the step function $\theta(\tau-t)$ in \eqref{b6}. With $\tau=\infty$ as in \cite{bon14,ter16}, the primary vessel keeps injecting tip density for all time. However, this may be artificial, as the primary vessel does not inject any more vessels after $t=0+$ in many experiments on early stage angiogenesis. Then $\tau$ in \eqref{b6} may be a small time of the order of the time step used in a numerical code. The original boundary condition in \cite{bon14} did not include the unit step function and, as a consequence, the deterministic description given by the tip density equation and its boundary conditions had an artificial injection of tip density at $x=0$ for all $t>0$. The deterministic description including boundary conditions can be proved to have a solution \cite{CDN16}.

\setcounter{equation}{0}
\renewcommand{\theequation}{B.\arabic{equation}}
\section{Derivation of equation \eqref{eq50} and meaning of the CCEs}\label{app3}
Let $\mathcal{I}=(a,b)$ and let us assume that $\xi+X=x/\epsilon$, with $\epsilon\ll 1$, for a fixed time. Let us consider the initial value problem
\begin{equation}
\frac{d\Lambda}{dx}=F(\tilde{p}(\xi;x),x),\quad \Lambda(a)=0,\label{c1}
\end{equation}
and solve it by using multiple scales $x$ and $\xi$, and the assumption $\Lambda=\Lambda^{(0)}(\xi,x)+\epsilon\Lambda^{(1)}(x,\xi)+O(\epsilon^2)$. We find the hierarchy of equations
\begin{eqnarray}
&&\frac{\partial\Lambda^{(0)}}{\partial\xi}=0,\label{c2}\\
&&\frac{\partial\Lambda^{(1)}}{\partial\xi}= F(\tilde{p}(\xi;x),x)-\frac{\partial\Lambda^{(0)}}{\partial x},\label{c3}
\end{eqnarray}
and so on. \eqref{c2} means that $\Lambda^{(0)}$ depends only on $x$. Assuming boundary conditions $\tilde{p}(\pm\mathcal{L}/2;x)=0$, \eqref{c3} has a solution bounded in $\xi$ for large $\xi$ provided the integral of its right hand side over $\xi$ is zero, which yields
\begin{equation}
\frac{\partial\Lambda^{(0)}}{\partial x}=\frac{1}{\mathcal{L}}\int_{-\mathcal{L}/2}^{\mathcal{L}/2}F(\tilde{p}(\xi;x),x)\, d\xi.\label{c4}
\end{equation}
Then $\Lambda^{(0)}(b)$ gives the formula \eqref{eq50}, which is typical in homogenization theory.

Consider now Eq. \eqref{eq46} with $\mu(C)=\int_a^b\mu(C(t,x,0))\, dx/(b-a)\equiv\overline{\mu}$ and a similar definition for $\overline{F_x}$. According to the assumptions specified below \eqref{eq46}, we may write
\begin{eqnarray}
&&\tilde{p}_s\!=\tilde{p}_s\!\left(\xi;K,c,\overline{\mu},F_x\!\left(C,\frac{\partial C}{\partial x}\right)\!\right)\!\!=\!\tilde{p}_s(\xi;K,c,\overline{\mu},\overline{F_x})\nonumber\\
&&\quad+\frac{\partial\tilde{p}_s}{\partial F_x}(\xi;K,c,\overline{\mu},\overline{F_x})(F_x-\overline{F_x})\!+\!\ldots.    \label{c5}
\end{eqnarray}
Then
\begin{eqnarray}
&&\nabla_x\tilde{p}_s=\mathbf{e}_{x}\frac{\partial\tilde{p}_s}{\partial\xi}(\xi;K,c,\overline{\mu},\overline{F_x})\nonumber\\
&&\quad+\frac{\partial\tilde{p}_s}{\partial F_x}(\xi;K,c,\overline{\mu},\overline{F_x})\nabla_xF_x+\ldots,
\end{eqnarray}
and similarly for $\Delta_x\tilde{p_s}$. Here $\mathbf{e}_x=(1,0)$. Using these formulas, we find $\mathcal{A}$ in \eqref{eq49} with the following meaning:
\begin{eqnarray}
&&\mathcal{A}=\! \frac{1}{2\beta}\frac{\partial^2\tilde{p}_s}{\partial\xi^2}(\xi;K,c,\overline{\mu},\overline{F_x})
-\tilde{p}_s(\xi;K,c,\overline{\mu},\overline{F_x})\nabla_x\!\cdot\!\mathbf{F}\nonumber\\&&\quad
-\frac{\partial\tilde{p}_s}{\partial F_x}(\xi;K,c,\overline{\mu},\overline{F_x})\!\left[\mathbf{F}\!\cdot\!\nabla_xF_x\!-\!\frac{1}{2\beta}\Delta_xF_x\right]\!  \nonumber\\&&\quad
+\frac{1}{\beta}\frac{\partial^2\tilde{p}_s}{\partial\xi\partial F_x}(\xi;K,c,\overline{\mu},\overline{F_x})\frac{\partial F_x}{\partial x}. \label{c6}
\end{eqnarray}
Then the CCEs become
\begin{eqnarray}
&&\dot{K}= \frac{(2K\Gamma\!+\!\overline{\mu}^2)^2}{4\Gamma\beta(c\!-\!\overline{F_x})^2}\frac{\frac{4\pi^2}{75}\!+\!\frac{1}{5}\!+\!\!\left(\frac{2\overline{F_x}}{5 c}\!-\!\frac{2\pi^2}{75}\!-\!\frac{9}{10}\right)\!\!\frac{\overline{F_x}}{c}}{\left(1-\frac{4\pi^2}{15}\right)\!\left(1-\frac{\overline{F_x}}{2c}\right)^2}  \nonumber\\&&\quad
-\frac{2K\Gamma\!+\!\overline{\mu}^2}{2\Gamma\!\left(c-\frac{\overline{F_x}}{2}\right)\!}\!\left(c\overline{\nabla_x\!\cdot\mathbf{F}}\! +\!\overline{\mathbf{F}\!\cdot\!\nabla_x F_x}-\frac{\overline{\Delta_xF_x}}{2\beta}\right)\!, \label{c7}
\end{eqnarray}
\begin{eqnarray}
\dot{c}=-\frac{7(2K\Gamma+\overline{\mu}^2)}{20\beta(c-\overline{F_x})}\frac{1-\frac{4\pi^2}{105}}{\left(1-\frac{4\pi^2}{15}\right)\!\left(1-\frac{\overline{F_x}}{2c}\right)\!}  \nonumber\\
+\frac{\overline{\mathbf{F}\!\cdot\!\nabla_x F_x} -(c-\overline{F_x})\overline{\nabla_x\!\cdot\!\mathbf{F}}-\frac{\overline{\Delta_xF_x}}{2\beta}}{2-\frac{\overline{F_x}}{c}},
\label{c8}
\end{eqnarray}
as indicated in Section \ref{sec:cc}.

\setcounter{equation}{0}
\renewcommand{\theequation}{C.\arabic{equation}}
\section{Extended collective coordinate equations for a soliton far from primary vessel and tumor}\label{app1}
In this Appendix, we will find the CCEs without the assumptions that $\mu$ is constant and that the time variation of the TAF concentration is negligible. To obtain the CCEs, we need to substitute the soliton \eqref{eq43} into \eqref{eq31}. According to \eqref{eq43}, the soliton is a function
\begin{equation}
\tilde{p}_s=\tilde{p}_s\!\left(\xi;K,c,\mu(C),F_x\!\left(C,\frac{\partial C}{\partial x}\right)\!\right)\!, \label{a1}
\end{equation}
so that we have the expressions:
\begin{widetext}
\begin{eqnarray}
&&\nabla_x\tilde{p}_s=\mathbf{e}_{x}\frac{\partial\tilde{p}_s}{\partial\xi}+\frac{\partial\tilde{p}_s}{\partial\mu}\mu_C\nabla_xC+ \frac{\partial\tilde{p}_s}{\partial F_x}\nabla_xF_x
=\mathbf{e}_{x}\frac{\partial\tilde{p}_s}{\partial\xi}+\frac{\partial\tilde{p}_s}{\partial K}\frac{\mu\mu_C}{\Gamma}\nabla_xC+ \frac{\partial\tilde{p}_s}{\partial F_x}\nabla_xF_x,\label{a2}\\
&&\Delta_x\tilde{p}_s=\frac{\partial^2\tilde{p}_s}{\partial\xi^2}+\frac{\partial\tilde{p}_s}{\partial K}\frac{\mu\mu_C\Delta_xC+(\mu_C^2+\mu\mu_{CC})|\nabla_xC|^2}{\Gamma}
+\frac{\partial\tilde{p}_s}{\partial F_x}\Delta_xF_x+\frac{\partial^2\tilde{p}_s}{\partial K^2} \frac{\mu^2\mu_C^2}{\Gamma^2}|\nabla_xC|^2\nonumber\\
&&\quad+\frac{\partial^2\tilde{p}_s}{\partial F_x^2} |\nabla_xF_x|^2 +2\frac{\mu\mu_C}{\Gamma}\!\left(\frac{\partial^2\tilde{p}_s}{\partial K\partial F_x} \nabla_xC\!\cdot\!\nabla_xF_x+\frac{\partial^2\tilde{p}_s}{\partial\xi\partial K}\frac{\partial C}{\partial x}\right)\!
+2\frac{\partial^2\tilde{p}_s}{\partial\xi\partial F_x}\frac{\partial F_x}{\partial x},\label{a3}\\
&&\frac{\partial\tilde{p}_s}{\partial t}=\frac{\partial\tilde{p}_s}{\partial K}\dot{K}+\frac{\partial\tilde{p}_s}{\partial c}\dot{c}-c\frac{\partial\tilde{p}_s}{\partial\xi} +\frac{\mu\mu_C}{\Gamma}\frac{\partial\tilde{p}_s}{\partial K}\frac{\partial C}{\partial t}
+\frac{\partial\tilde{p}_s}{\partial F_x} \frac{\partial F_x}{\partial t}, \label{a4}\\
&&\frac{\partial F_x}{\partial t}=\frac{\delta}{\beta}\frac{\partial}{\partial t}\frac{\frac{\partial C}{\partial x}}{(1+\Gamma_1C)^q}=\frac{\delta}{\beta(1+\Gamma_1C)^q}\!\left(\frac{\partial^2C}{\partial t\partial x}
-\frac{\partial C}{\partial x}\frac{\partial C}{\partial t}\frac{q\Gamma_1}{1+\Gamma_1C}\right)\!
=\frac{\delta}{\beta}\frac{\partial}{\partial x}\frac{\frac{\partial C}{\partial t}}{(1+\Gamma_1C)^q}, \label{a5}
\end{eqnarray}
\end{widetext}
in which $\mathbf{e}_{x}$ is the unit vector along the $x$ axis and we have used
\begin{equation}
\frac{\partial\tilde{p}_s}{\partial\mu}=\frac{\partial\tilde{p}_s}{\partial K}\frac{\mu}{\Gamma}. \label{a6}
\end{equation}
Inserting \eqref{a2}-\eqref{a5} in \eqref{eq31}, we obtain 
\begin{eqnarray}
&&\left(F_x-\dot{X}\right)\!\frac{\partial\tilde{p}_s}{\partial\xi}+\frac{\partial\tilde{p}_s}{\partial K} \dot{K}+\frac{\partial\tilde{p}_s}{\partial c}\dot{c}+\tilde{p}_s\nabla_x\cdot\mathbf{F}  \nonumber\\&&
+\frac{\partial\tilde{p}_s}{\partial\mu}\mu_{C}\!\left(\frac{\partial C}{\partial t}+\mathbf{F}\cdot\nabla_xC\right)\!+\frac{\partial\tilde{p}_s}{\partial F_x}\!\left(\frac{\partial F_x}{\partial t}\right.\nonumber\\
&&\left.+\mathbf{F}\cdot\nabla_xF_x\right)\! -\frac{1}{2\beta}\!\left(\frac{\partial^2\tilde{p}_s}{\partial\xi^2} +2\frac{\partial^2\tilde{p}_s}{\partial\xi\partial\mu}\mu_C\frac{\partial C}{\partial x} \right.\nonumber\\&&
+2\frac{\partial^2\tilde{p}_s}{\partial\xi\partial F_x}\frac{\partial F_x}{\partial x}+ \frac{\partial\tilde{p}_s}{\partial F_x}\Delta_xF_x+\frac{\partial^2\tilde{p}_s}{\partial F_x^2}|\nabla_xF_x|^2 \nonumber\\
&&+\frac{\partial\tilde{p}_s}{\partial K}\frac{\mu\mu_C\Delta_xC+(\mu\mu_{CC}+\mu_C^2)|\nabla_xC|^2}{\Gamma}  \nonumber\\&&
\left.+\frac{\partial^2\tilde{p}_s}{\partial K^2}\frac{\mu^2\mu_C^2}{\Gamma^2} |\nabla_xC|^2+ 2 \frac{\partial^2\tilde{p}_s}{\partial K\partial F_x}\frac{\mu\mu_C}{\Gamma} \nabla_xC\!\cdot\!\nabla_x F_x\right)\!\nonumber\\
&&=\mu\tilde{p}_s-\Gamma\tilde{p}_s\!\!\int_0^t\tilde{p}_s dt. \label{a7}
\end{eqnarray}
Substituting \eqref{eq48} (with $\tilde{p}=\tilde{p}_s$) in \eqref{a5}, we obtain 
\begin{eqnarray}
\frac{\partial F_x}{\partial t}=\frac{\kappa\delta}{\beta}\frac{\partial}{\partial x}\frac{\Delta_xC}{(1+\Gamma_1C)^q} - \frac{\chi\delta}{\beta}\tilde{p}_s\frac{\partial}{\partial x}\frac{C}{(1+\Gamma_1C)^q} \nonumber\\ 
- \frac{\chi\delta C}{\beta(1+\Gamma_1C)^q}\!\left(\frac{\partial\tilde{p}_s}{\partial\xi}+\frac{\mu\mu_C}{\Gamma}\frac{\partial\tilde{p}_s}{\partial K}\frac{\partial C}{\partial x}+\frac{\partial\tilde{p}_s}{\partial F_x} \frac{\partial F_x}{\partial x}\right)\!. \label{a8}
\end{eqnarray}
Inserting these equations into \eqref{a7} and using \eqref{eq43} and \eqref{eq44}, we obtain \eqref{eq33} with the following $\mathcal{A}$,
\begin{widetext}
\begin{eqnarray}
&&\mathcal{A}=  \frac{1}{2\beta}\frac{\partial^2\tilde{p}_s}{\partial\xi^2}-\tilde{p}_s\nabla_x\cdot\mathbf{F}-\frac{\partial\tilde{p}_s}{\partial K}\!\left[\frac{\mu\mu_C}{\Gamma}\!\left(\mathbf{F}\cdot\nabla_xC+\kappa\Delta_xC-\chi C\tilde{p}_s-\frac{1}{2\beta}\Delta_xC\right)\! \right. \nonumber\\
&&\quad \left.-\frac{\mu_C^2+\mu\mu_{CC}}{2\beta\Gamma}|\nabla_xC|^2\right]\! 
-\frac{\partial\tilde{p}_s}{\partial F_x}\!\left[\mathbf{F}\cdot\nabla_xF_x+\frac{\kappa\delta}{\beta}\frac{\partial}{\partial x}\!\left(\frac{\Delta_xC}{(1+\Gamma_1C)^q}\right)\!-\frac{1}{2\beta}\Delta_xF_x\right]\!\nonumber\\
&&\quad +\frac{\mu^2\mu_C^2}{2\beta\Gamma^2}\frac{\partial^2\tilde{p}_s}{\partial K^2}|\nabla_xC|^2+\frac{\mu\mu_C}{\beta\Gamma}\frac{\partial^2\tilde{p}_s}{\partial K\partial F_x}\nabla_xC\!\cdot\!\nabla_xF_x +\frac{\delta\chi}{\beta}\frac{\partial}{\partial x}\!\left(\frac{C}{(1+\Gamma_1C)^q}\right)\!\frac{\partial\tilde{p}_s}{\partial F_x}\tilde{p}_s\nonumber\\&&\quad 
+\frac{1}{2\beta}\frac{\partial^2\tilde{p}_s}{\partial F_x^2}|\nabla_xF_x|^2
+ \frac{\delta\chi C}{\beta(1+\Gamma_1C)^q}\frac{\partial F_x}{\partial x}\!\left(\frac{\partial\tilde{p}_s}{\partial F_x}\right)^2\!+\frac{\delta\chi\mu\mu_C C}{\beta\Gamma(1+\Gamma_1C)^q}\frac{\partial C}{\partial x}\frac{\partial\tilde{p}_s}{\partial K}\frac{\partial\tilde{p}_s}{\partial F_x} \nonumber\\
&&\quad+\frac{\mu\mu_C}{\Gamma\beta}\frac{\partial^2\tilde{p}_s}{\partial\xi\partial K}\frac{\partial C}{\partial x}+\frac{1}{\beta}\frac{\partial^2\tilde{p}_s}{\partial\xi\partial F_x}\frac{\partial F_x}{\partial x}+\frac{\delta\chi C}{\beta(1+\Gamma_1C)^q}\frac{\partial\tilde{p}_s}{\partial F_x}\frac{\partial\tilde{p}_s}{\partial\xi}, \label{a9}
\end{eqnarray}
\end{widetext}
instead of \eqref{eq49}.

After we calculate the integrals that appear in (\ref{eq51})-\eqref{eq52} as indicated in Appendix \ref{app4}, these equations become:
\begin{widetext}
\begin{eqnarray}
&&\dot{K}= \frac{(2K\Gamma+\mu^2)^2}{4\Gamma\beta(c-F_x)^2}\frac{\frac{4\pi^2}{75}+\frac{1}{5}+\frac{F_x}{c}\!\left(\frac{2F_x}{5 c}-\frac{2\pi^2}{75}-\frac{9}{10}\right)\!}{\left(1-\frac{4\pi^2}{15}\right)\!\left(1-\frac{F_x}{2c}\right)^2}-\frac{2K\Gamma+\mu^2}{2\Gamma c\!\left(1-\frac{F_x}{2c}\right)}\!\left[c\nabla_x\!\cdot\!\mathbf{F} +\mathbf{F}\!\cdot\!\nabla_x F_x -\frac{1}{2\beta}\Delta_xF_x\right.\nonumber\\
&&\quad\left. +\frac{\kappa\delta}{\beta}\frac{\partial}{\partial x}\!\left(\frac{\Delta_xC}{(1+\Gamma_1C)^q}\right)\!\right]\! -\frac{\mu\mu_C}{\Gamma}\!\left[\mathbf{F}\!\cdot\!\nabla_x C+\!\left(\kappa-\frac{1}{2\beta}\right)\!\Delta_xC-\frac{\chi c C(2K\Gamma+\mu^2)\!\left[1-\frac{34\pi^2}{105}\!\left(1-\frac{6F_x}{17c}\right)\right]\!}{3\Gamma(c-F_x)\left(1-\frac{4\pi^2}{15}\right)\!\left(1-\frac{F_x}{2c}\right)\!}\right]\nonumber\\
&&\quad+\frac{\chi\delta c(2K\Gamma+\mu^2)^2}{5\beta\Gamma^2(c-F_x)^2\!\left(1-\frac{4\pi^2}{15}\right)\!\left(1-\frac{F_x}{2c}\right)\!}\!\left\{\frac{C}{6(1+\Gamma_1C)^q}\!\left[\frac{5\mu\mu_C\frac{\partial C}{\partial x}\!\left[1-\frac{2\pi^2}{105}\!\left(17+\frac{4\pi^2}{5}\right)\!+\!\frac{8\pi^2F_x}{35c}\right]\!}{2K\Gamma+\mu^2}\right.\right.\nonumber\\
&&\quad\left.\left.-\frac{\partial F_x}{\partial x} \frac{1-\frac{2\pi^2}{105}+\frac{8\pi^4}{105}-\frac{4F_x}{c}}{c-F_x}\!\right]\!+\frac{\partial}{\partial x}\!\left(\frac{ C}{(1+\Gamma_1C)^q}\right)\!\left[1-\frac{86\pi^2}{315}-\frac{2F_x}{3c}\!\left(1-\frac{2\pi^2}{7}\right)\!\right]\!\right\}\!\nonumber\\
&&\quad +\frac{|\nabla_xC|^2}{2\beta\Gamma}\!\left[\mu_C^2+\mu\mu_{CC}\! 
-\!\frac{\mu^2\mu_C^2\!\left(\frac{\pi^2}{15}+\frac{F_x}{c}\right)}{2(2K\Gamma+\mu^2)\!\left(1-\frac{F_x}{2c}\right)\!}\right]\!+\frac{\!1\!-\!\frac{\pi^2}{30}\!-\!\frac{3F_x}{2c}\!\left(1\!-\!\frac{\pi^2}{90}\!\right)\!+\!\frac{F_x^2}{2c^2}\!}{\beta\Gamma(c-F_x)\!\left(1-\frac{F_x}{2c}\right)^2\!} (2K\Gamma+\mu^2)\nonumber\\
&&\quad \times\!\left[\frac{\mu\mu_C\nabla_xC\!\cdot\!\nabla_xF_x}{2K\Gamma+\mu^2} +\frac{|\nabla_xF_x|^2}{2(c-F_x)}\right]\! ,  \label{a10}
\end{eqnarray}
\begin{eqnarray}
&&\dot{c}=\!-\frac{7(2K\Gamma+\mu^2)\!\left(1-\frac{4\pi^2}{105}\right)\!}{20\beta(c\!-\!F_x)\!\left(1\!-\!\frac{4\pi^2}{15}\right)\!\left(1\!-\!\frac{F_x}{2c}\right)\!}\!+\!\frac{\mathbf{F}\!\cdot\!\nabla_x F_x\! +\!\frac{\kappa\delta}{\beta}\frac{\partial}{\partial x}\!\left(\frac{\Delta_xC}{(1+\Gamma_1C)^q}\right)\!-\!\frac{\Delta_xF_x}{2\beta}\!-\!(c-F_x)\nabla_x\!\cdot\!\mathbf{F}}{2-\frac{F_x}{c}}\nonumber\\
&&\quad -\!\left[\frac{\mu\mu_C\nabla_xC\!\cdot\!\nabla_xF_x}{2K\Gamma+\mu^2} 
+\frac{|\nabla_xF_x|^2}{2(c-F_x)}\right]\! \frac{1+\frac{\pi^2}{30}}{\beta \!\left(1\!-\!\frac{F_x}{2c}\right)\!}+\frac{\chi\delta c (2K\Gamma+\mu^2)}{3\beta\Gamma\!\left(1-\frac{4\pi^2}{15}\right)\!\left(1-\frac{F_x}{2c}\right)\!}\!\left\{\frac{\mu\mu_CC}{2K\Gamma+\mu^2} \right.\nonumber\\
&&\quad \times\!\left[\frac{1+\frac{2\pi^2}{15}\!\left(1-\frac{4\pi^2}{35}\right)\!}{2(c-F_x)(1+\Gamma_1C)^q}\frac{\partial C}{\partial x}\!+\!\frac{\beta}{\delta}\!\left(1\!-\!\frac{2\pi^2}{21}\right)\!\right]
\!+\frac{7C\!\left[1\!+\!\frac{2\pi^2(1\!-\!4\pi^2)}{735}\right]\!}{10(1\!+\!\Gamma_1C)^q(c\!-\!F_x)^2\!}\frac{\partial F_x}{\partial x}\nonumber\\
&&\quad -\frac{1\!-\!\frac{34\pi^2}{105}}{5(c-F_x)}\!\left.\frac{\partial}{\partial x}\!\!\left(\frac{ C}{(1+\Gamma_1C)^q}\!\right)\!\right\} \! 
- \frac{\mu^2\mu_{C}^2(c-F_x)}{2\beta(2K\Gamma+\mu^2)^2\!\left(1-\frac{F_x}{2c}\right)\!}\!\left(1\!+\!\frac{\pi^2}{30}\right)\!|\nabla_xC|^2\! .
\label{a11}
\end{eqnarray}
\end{widetext}

In Table \ref{table2}, the dimensionless coefficients $\kappa$ and $\chi$ appearing in the TAF equation \eqref{eq33} are very small. Then we may ignore terms having these coefficients in the CCEs \eqref{a10}-\eqref{a11}, thereby obtaining
\begin{widetext}
\begin{eqnarray}
&&\dot{K}= \frac{(2K\Gamma+\mu^2)^2}{4\Gamma\beta(c-F_x)^2}\frac{\frac{4\pi^2}{75}+\frac{1}{5}+\frac{F_x}{c}\!\left(\frac{2F_x}{5 c}-\frac{2\pi^2}{75}-\frac{9}{10}\right)\!}{\left(1-\frac{4\pi^2}{15}\right)\!\left(1-\frac{F_x}{2c}\right)^2}-\frac{2K\Gamma+\mu^2}{2\Gamma\!\left(c-\frac{F_x}{2}\right)}\!\left(c\nabla_x\!\cdot\!\mathbf{F} +\mathbf{F}\!\cdot\!\nabla_x F_x -\frac{1}{2\beta}\Delta_xF_x\right)\!\nonumber\\
&&\quad -\frac{\mu\mu_C}{\Gamma}\!\left(\mathbf{F}\!\cdot\!\nabla_x C-\frac{\Delta_xC}{2\beta}\right)\!+\!\left[\frac{\mu_C^2+\mu\mu_{CC}}{\Gamma}-\frac{\mu^2\mu_C^2\!\left(\frac{\pi^2}{15}+\frac{F_x}{c}\right)}{2\Gamma(2K\Gamma+\mu^2)\!\left(1-\frac{F_x}{2c}\right)\!}\right]\!\frac{|\nabla_xC|^2}{2\beta}\nonumber\\
&&\quad +\frac{\!1\!-\!\frac{\pi^2}{30}\!-\!\frac{3F_x}{2c}\!\left(1\!-\!\frac{\pi^2}{90}\!\right)\!+\!\frac{F_x^2}{2c^2}\!}{\beta\Gamma(c-F_x)\!\left(1-\frac{F_x}{2c}\right)^2\!} (2K\Gamma+\mu^2)\!\left[\frac{\mu\mu_C\nabla_xC\!\cdot\!\nabla_xF_x}{2K\Gamma+\mu^2} +\frac{|\nabla_xF_x|^2}{2(c-F_x)}\right]\! ,  \label{a12}
\end{eqnarray}
\begin{eqnarray}
&&\dot{c}=-\frac{7(2K\Gamma+\mu^2)}{20\beta(c-F_x)}\frac{1-\frac{4\pi^2}{105}}{\left(1-\frac{4\pi^2}{15}\right)\!\left(1-\frac{F_x}{2c}\right)\!}+\frac{\mathbf{F}\!\cdot\!\nabla_x F_x -\frac{\Delta_xF_x}{2\beta}-(c-F_x)\nabla_x\!\cdot\!\mathbf{F}}{2-\frac{F_x}{c}}\nonumber\\
&&\quad -\!\left[\frac{\mu^2\mu_C^2(c-F_x)|\nabla_xC|^2}{2(2K\Gamma+\mu^2)^2}+\frac{\mu\mu_C\nabla_xC\!\cdot\!\nabla_xF_x}{2K\Gamma+\mu^2} +\frac{|\nabla_xF_x|^2}{2(c-F_x)}\right]\! \frac{1+\frac{\pi^2}{30}}{\beta \!\left(1-\frac{F_x}{2c}\right)\!} .
\label{a13}
\end{eqnarray}
\end{widetext}
Further simplification leads to \eqref{eq53}-\eqref{eq54}.

\begin{figure}[h]
\begin{center}
\includegraphics[width=9.0cm]{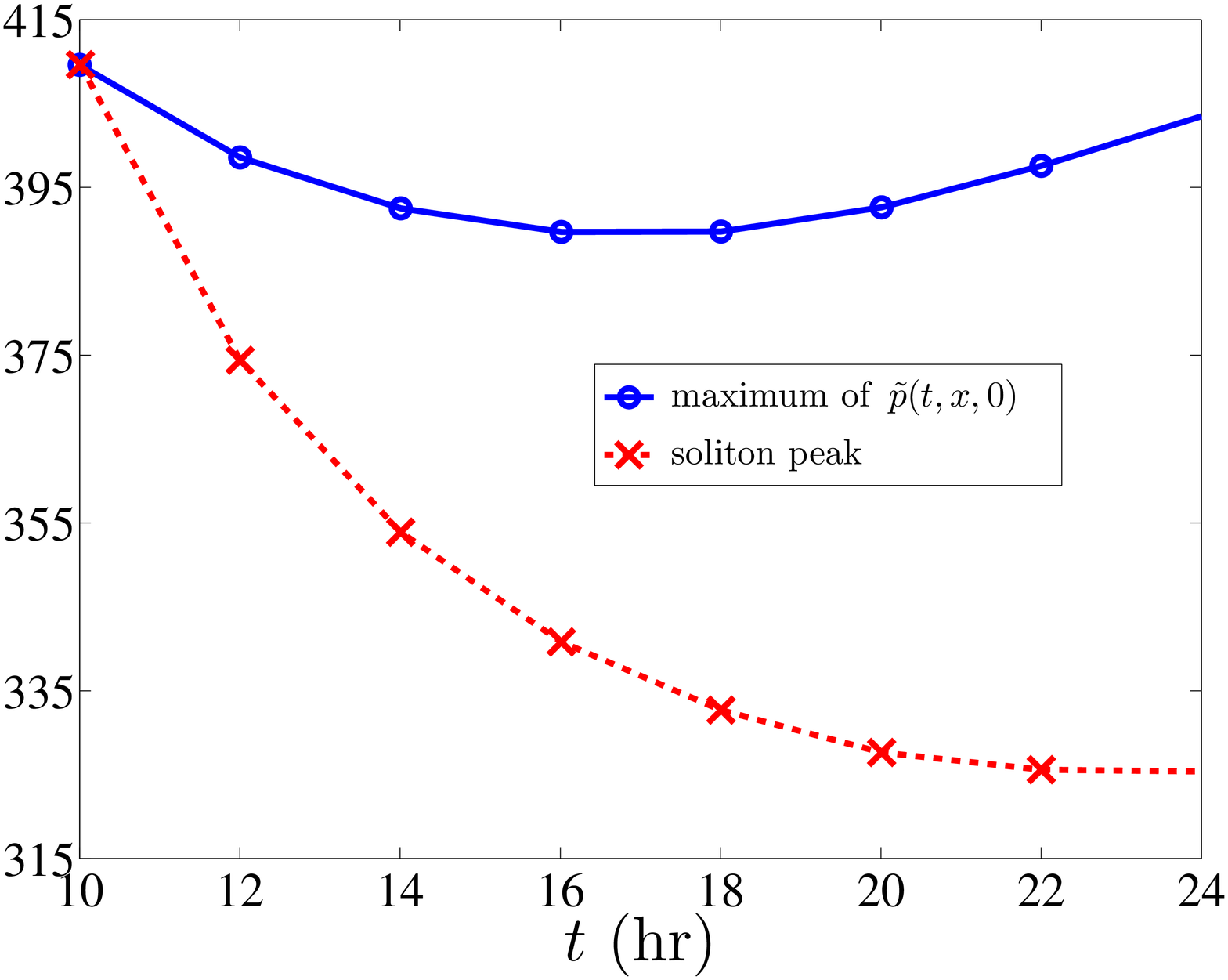} 
\end{center}
\vskip-4mm \caption{Same as Figure \ref{fig2}: Comparison between the maximum value of  $\tilde{p}(t,x,0)$ as given by the deterministic description and its value as predicted by soliton collective coordinates that solve \eqref{a10}-\eqref{a11}. \label{fig8}}
\end{figure}

We can reconstruct the soliton using the extended CCEs \eqref{a10}-\eqref{a11} instead of the simplified CCEs \eqref{eq53}-\eqref{eq54}. Somewhat surprisingly, the reconstruction compares  poorly with the direct solution of the deterministic description. Figure \ref{fig8} depicts the evolution of the soliton peak when evaluated from \eqref{a10}-\eqref{a11} and the peak of the reduced density $\tilde{p}(t,x,0)$ as given by the deterministic description. We observe that the soliton peak decreases far away from $\tilde{p}(t,x,0)$. The reason is that $K(t)$ monotonically decreases with $t$. Instead, $K(t)$ given by \eqref{eq53}-\eqref{eq54} reaches a minimum and it increases as shown in Figure \ref{fig1}(a). Then the soliton peak calculated from the CCEs \eqref{eq53}-\eqref{eq54} and depicted in Figure \ref{fig2}(a) increases after reaching a local minimum and it becomes closer to $\tilde{p}(t,x,0)$. The discrepancies between the solutions of the different CCEs are caused by the terms proportional to $\mu\mu_C$ in \eqref{a10}. In particular, the negative term $-\frac{\mu\mu_C}{\Gamma}\!\left(\mathbf{F}\!\cdot\!\nabla_x C-\frac{\Delta_xC}{2\beta}\right)\!$ in \eqref{a10} or in \eqref{a12} is too large to be compensated by any positive term in the equation for $\dot{K}$. In turn, the large value of $A$ in Table \ref{table2} amplifies the importance of the spatial variation of $C$, $\mathbf{F}\!\cdot\!\nabla_x C$, reflected in that coefficient. In principle, the CCEs are based on the idea that the spatial variations of $C$, which appear when $\tilde{p}_s$ of \eqref{eq43} and \eqref{eq46} is differentiated with respect to $x$, produce terms that are small compared to $\partial\tilde{p}_s/\partial\xi$. The large value of $A$ contradicts this idea and thus the CCEs \eqref{eq53}-\eqref{eq54} based on setting $\mu_C=0$ give better results than \eqref{a10}-\eqref{a11} or \eqref{a12}-\eqref{a13}.

\setcounter{equation}{0}
\renewcommand{\theequation}{D.\arabic{equation}}
\section{Derivation of the collective coordinate equations}\label{app4}
The derivatives of $\tilde{p}_s$, given by \eqref{eq43}-\eqref{eq45}, which appear in (\ref{a9}) are:
\begin{eqnarray}
&&\frac{\partial^2\tilde{p}_s}{\partial\xi^2}=\frac{c(2K\Gamma+\mu^2)^{2}}{4\Gamma(c-F_x)^3}\mbox{sech}^4 s\, (2\sinh^2s-1),  \label{d1}\\
&&\frac{\partial\tilde{p}_s}{\partial K}=\frac{c}{c-F_x}\mbox{sech}^2 s\, (1-s\tanh s),\label{d2}\\
&&\frac{\partial\tilde{p}_s}{\partial c}=\frac{c(2K\Gamma+\mu^2)}{\Gamma(c-F_x)^2}\mbox{sech}^2 s\!\left(s\tanh s-\frac{F_x}{2 c}\right)\!\!, \label{d3}\\
&&\frac{\partial\tilde{p}_s}{\partial\mu}=\frac{\mu}{\Gamma}\frac{\partial\tilde{p}_s}{\partial K},\label{d4}\\
&&\frac{\partial\tilde{p}_s}{\partial F_x}=\frac{c(2K\Gamma+\mu^2)}{\Gamma(c-F_x)^2}\mbox{sech}^2 s\left(\frac{1}{2}-s\tanh s\right)\!\!, \label{d5}\\
&&\frac{\partial^2\tilde{p}_s}{\partial K^2}=\frac{c\Gamma\, s\frac{\partial}{\partial s}[\mbox{sech}^2 s\, (1-s\tanh s)]}{(c-F_x)(2K\Gamma+\mu^2)},  \label{d6} \\
&&\frac{\partial^2\tilde{p}_s}{\partial K\partial F_x}=\frac{c\, s}{(c-F_x)^2}\frac{\partial}{\partial s}[\mbox{sech}^2 s\, (1-s\tanh s)]\nonumber\\
&&\quad\quad\quad\quad +\frac{1}{c-F_x}\, \frac{\partial\tilde{p}_s}{\partial K}, \label{d7}\\
&&\frac{\partial^2\tilde{p}_s}{\partial F_x^2}=\frac{2}{c-F_x}\frac{\partial\tilde{p}_s}{\partial F_x} +\frac{c(2K\Gamma+\mu^2)}{\Gamma(c-F_x)^3}\, s\nonumber\\
&&\quad\quad\,\,\,\times\, \frac{\partial}{\partial s}\!\left[\mbox{sech}^2 s\!\left(\frac{1}{2}-s\tanh s\right)\!\right]\!, \label{d8}
\end{eqnarray}
and
\begin{eqnarray}
\mu_C\!\equiv \frac{\partial\mu}{\partial C}=\frac{d}{\pi(1+C)^2}\!\!\left[1+\frac{\alpha\,\ln\!\left(1+\frac{1}{\sigma_v^2}\right)\!}{\pi\beta(1+\sigma_v^2)}\right]\!\!. \label{d9}
\end{eqnarray}
To find the CCEs of the soliton, we need the following integrals calculated from \eqref{d1}-\eqref{d9}:
\begin{eqnarray}
\int_{-\infty}^\infty\!\!\left(\frac{\partial\tilde{p}_s}{\partial K}\right)^2\!\!d\xi\!=\!\frac{4c^2(2K\Gamma+\mu^2)^{-\frac{1}{2}}}{3(c-F_x)}\!\!\left(\!1+\frac{\pi^2}{30}\!\right)\!,  \label{d10}
\end{eqnarray}
\begin{eqnarray}
&&\int_{-\infty}^\infty\!\!\frac{\partial\tilde{p}_s}{\partial K}\frac{\partial\tilde{p}_s}{\partial c}\!d\xi\!=\!\frac{2c^2(2K\Gamma+\mu^2)^{1/2}}{3\Gamma(c-F_x)^2}\nonumber\\
&&\quad\quad\quad\quad\quad\quad\quad\times
\left(1-\frac{3F_x}{2c} -\frac{\pi^2}{15}\right)\!,  \label{d11}
\end{eqnarray}
\begin{eqnarray}
&&\int_{-\infty}^\infty\!\!\left(\frac{\partial\tilde{p}_s}{\partial c}\right)^2\!\!d\xi\!=\!\frac{2c^2(2K\Gamma+\mu^2)^{\frac{3}{2}}}{3\Gamma^2(c-F_x)^3}\nonumber\\
&&\quad\quad\quad\quad\quad\quad\quad\times
\left(\frac{\pi^2}{15}\!-\!\frac{F_x(c-F_x)}{c^2}\!\right)\!, \label{d12}\\
&&\int_{-\infty}^\infty\!\!\frac{\partial\tilde{p}_s}{\partial K}\frac{\partial^2\tilde{p}_s}{\partial\xi^2}d\xi =-\frac{c^2(2K\Gamma+\mu^2)^{3/2}}{3\Gamma(c-F_x)^3}, \label{d13}
\end{eqnarray}
\begin{eqnarray}
&&\int_{-\infty}^\infty\!\!\frac{\partial\tilde{p}_s}{\partial c}\frac{\partial^2\tilde{p}_s}{\partial\xi^2}d\xi =\frac{c^2(2K\Gamma+\mu^2)^{5/2}}{15\Gamma^2(c-F_x)^4}\nonumber\\&&\quad\quad\quad\quad\quad\quad\quad\,\,\times
\left(1+\frac{2}{c}F_x\right)\!, \label{d14}\\
&&\int_{-\infty}^\infty\!\!\frac{\partial\tilde{p}_s}{\partial K}\tilde{p}_sd\xi =\frac{c^2(2K\Gamma+\mu^2)^{1/2}}{\Gamma(c-F_x)}, \label{d15}\end{eqnarray}
\begin{eqnarray}
&&\int_{-\infty}^\infty\!\!\frac{\partial\tilde{p}_s}{\partial c}\tilde{p}_s d\xi \!=\!\frac{c^2(2K\Gamma+\mu^2)^{3/2}}{3\Gamma^2(c-F_x)^2}
\!\!\left(\!1-\frac{2F_x}{c}\!\right)\!\!, \label{d16}\\
&&\int_{-\infty}^\infty\!\tilde{p}_s\!\left(\frac{\partial\tilde{p}_s}{\partial K}\right)^2 d\xi =\frac{2c^3(2K\Gamma+\mu^2)^{1/2}}{3\Gamma (c-F_x)^2}
\nonumber\\&&\quad\quad\quad\quad\quad\quad\quad\quad\,\times
\left(1+\frac{2\pi^2}{105}\right)\!\!, \label{d17}
\end{eqnarray}
\begin{eqnarray}
&&\int_{-\infty}^\infty\!\!\tilde{p}_s\frac{\partial\tilde{p}_s}{\partial c}\frac{\partial\tilde{p}_s}{\partial K} d\xi =\frac{2 c^3(2K\Gamma+\mu^2)^{3/2}}{9\Gamma^2(c-F_x)^3}\nonumber\\&&\quad\quad\quad\quad\quad\quad\quad\quad\times
\left(1-\frac{2\pi^2}{35}-\frac{2F_x}{c}\right)\!, \label{d18}\\
&&\int_{-\infty}^\infty\!\!\frac{\partial\tilde{p}_s}{\partial K}\frac{\partial\tilde{p}_s}{\partial F_x}d\xi =\frac{c^2(2K\Gamma+\mu^2)^{1/2}}{3\Gamma(c-F_x)^2}\nonumber\\&&\quad\quad\quad\quad\quad\quad\quad\times
\!\left(1+\frac{2\pi^2}{15}\right)\!, \label{d19}\\
&&\int_{-\infty}^\infty\!\!\frac{\partial\tilde{p}_s}{\partial c}\frac{\partial\tilde{p}_s}{\partial F_x}d\xi =\frac{c^2(2K\Gamma+\mu^2)^{3/2}}{3\Gamma^2(c-F_x)^3}\nonumber\\&&\quad\quad\quad\quad\quad\quad\quad\times
\!\left(1-\frac{2\pi^2}{15}-\frac{F_x}{c}\right)\!, \label{d20}
\end{eqnarray}
\begin{eqnarray}
&&\int_{-\infty}^\infty\!\!\tilde{p}_s\frac{\partial\tilde{p}_s}{\partial K}\frac{\partial\tilde{p}_s}{\partial F_x}d\xi =\frac{2c^3(2K\Gamma+\mu^2)^{3/2}}{9\Gamma^2(c-F_x)^3}\nonumber\\&&\quad\quad\quad\quad\quad\quad\quad\quad\times
\!\left(1+\frac{2\pi^2}{35}\right)\!, \label{d21}\\
&&\int_{-\infty}^\infty\!\!\tilde{p}_s\frac{\partial\tilde{p}_s}{\partial c}\frac{\partial\tilde{p}_s}{\partial F_x}d\xi =\frac{2c^3(2K\Gamma+\mu^2)^{5/2}}{15\Gamma^3(c-F_x)^4}\nonumber\\&&\quad\quad\quad\quad\quad\quad\quad\quad\times
\left(1-\frac{4F_x}{3c}-\frac{2\pi^2}{21}\right)\!,  \label{d22}
\end{eqnarray}
\begin{eqnarray}
&&\int_{-\infty}^\infty\!\!\left(\frac{\partial\tilde{p}_s}{\partial K}\right)^2\frac{\partial\tilde{p}_s}{\partial F_x}d\xi =\frac{c^3(2K\Gamma+\mu^2)^{1/2}}{3\Gamma (c-F_x)^3}\nonumber\\
&&\quad\quad\quad\quad\quad\quad\quad\quad\quad\times
\!\left(1+\frac{2\pi^2}{21}\right)\!,\label{d23}\\
&&\int_{-\infty}^\infty\!\!\frac{\partial\tilde{p}_s}{\partial K}\!\left(\frac{\partial\tilde{p}_s}{\partial F_x}\right)^2\!d\xi =\frac{c^3(2K\Gamma+\mu^2)^{3/2}}{9\Gamma^2 (c-F_x)^4}\nonumber\\&&\quad\quad\quad\quad\quad\quad\quad\quad\quad\times
\!\left(1+\frac{6\pi^2}{35}\right)\!, \label{d24}
\end{eqnarray}

\begin{eqnarray}
\int_{-\infty}^\infty\!\!\frac{\partial\tilde{p}_s}{\partial c}\!\left(\frac{\partial\tilde{p}_s}{\partial F_x}\right)^2 d\xi =\frac{c^3(2K\Gamma+\mu^2)^{5/2}}{45\Gamma^3(c-F_x)^5}\nonumber\\
\times
\!\left[\frac{2\pi^2}{7}-1-\frac{2F_x}{c}\!\left(1+\frac{2\pi^2}{7}\right)\!\right]\!, \label{d25}\\
\int_{-\infty}^\infty\!\!\frac{\partial\tilde{p}_s}{\partial K}\frac{\partial\tilde{p}_s}{\partial c}\frac{\partial\tilde{p}_s}{\partial F_x}d\xi =\frac{c^3(2K\Gamma+\mu^2)^{3/2}}{9\Gamma^2(c-F_x)^4}\nonumber\\
\times
\!\left[1-\frac{2\pi^2}{35}-\frac{2F_x}{c}\!\left(1+\frac{2\pi^2}{35}\right)\!\right]\!,  \label{d26}
\end{eqnarray}

\begin{eqnarray}
\int_{-\infty}^\infty\!\!\frac{\partial\tilde{p}_s}{\partial K} \frac{\partial^2\tilde{p}_s}{\partial K^2} d\xi =-\frac{2c^2\Gamma}{3(c-F_x)(2K\Gamma+\mu^2)^{3/2}}\nonumber\\
\times\left(1+\frac{\pi^2}{30}\right)\!, \label{d27}\end{eqnarray}
\begin{eqnarray}
\int_{-\infty}^\infty\!\!\frac{\partial\tilde{p}_s}{\partial c}\frac{\partial^2\tilde{p}_s}{\partial K^2}d\xi =\frac{c^2(2K\Gamma+\mu^2)^{-1/2}}{3(c-F_x)^2}\nonumber\\
\times
\!\left[-\frac{\pi^2}{5}+\frac{F_x}{c}\!\left(1+\frac{2\pi^2}{15}\right)\!\right]\!,  \label{d28}
\end{eqnarray}

\begin{eqnarray}
\int_{-\infty}^\infty\!\!\frac{\partial\tilde{p}_s}{\partial K} \frac{\partial^2\tilde{p}_s}{\partial K\partial F_x} d\xi =\frac{2c^2(2K\Gamma+\mu^2)^{-1/2}}{3(c-F_x)^2}\nonumber\\
\times\left(1+\frac{\pi^2}{30}\right)\!, \label{d29}\end{eqnarray}
\begin{eqnarray}
\int_{-\infty}^\infty\!\!\frac{\partial\tilde{p}_s}{\partial c}\frac{\partial^2\tilde{p}_s}{\partial K\partial F_x}d\xi =\frac{2c^2(2K\Gamma+\mu^2)^{1/2}}{3\Gamma(c-F_x)^3}\nonumber\\
\times
\!\left[1-\frac{\pi^2}{6}-\frac{F_x}{c}\!\left(1-\frac{\pi^2}{15}\right)\!\right]\!,  \label{d30}
\end{eqnarray}

\begin{eqnarray}
\!\int_{-\infty}^\infty\!\!\frac{\partial\tilde{p}_s}{\partial K} \frac{\partial^2\tilde{p}_s}{\partial F_x^2} d\xi\!=\!\frac{2c^2(2K\Gamma+\mu^2)^{1/2}}{3\Gamma(c-F_x)^3}\!\left(1+\frac{\pi^2}{30}\right)\!\!, \label{d31}\end{eqnarray}
\begin{eqnarray}
\int_{-\infty}^\infty\!\!\frac{\partial\tilde{p}_s}{\partial c}\frac{\partial^2\tilde{p}_s}{\partial F_x^2}d\xi =\frac{2c^2(2K\Gamma+\mu^2)^{3/2}}{3\Gamma^2(c-F_x)^4}\nonumber\\
\times
\!\left[1-\frac{\pi^2}{6}-\frac{F_x}{c}\!\left(1-\frac{\pi^2}{15}\right)\!\right]\!.  \label{d32}
\end{eqnarray}
Using these integrals, we obtain the CCEs (\ref{a10}) and (\ref{a11}).


\begin{thebibliography}{99}
\bibitem{fol74} J. Folkman, Tumor angiogenesis. Adv. Cancer Res. {\bf 19}, 331-358 (1974).
\bibitem{car05} P. F.  Carmeliet, Angiogenesis in life, disease and medicine. Nature {\bf 438}, 932-936 (2005).
\bibitem{car11} P. Carmeliet, and R.K. Jain, Molecular mechanisms and clinical applications of angiogenesis. Nature {\bf 473}, 298-307 (2011).
\bibitem{fig08} W.D. Figg, and J. Folkman (eds), \emph{Angiogenesis. An Integrative Approach From Science to Medicine} (Springer, Berlin 2008).
\bibitem{CT05} P. Carmeliet, and M. Tessier-Lavigne, Common mechanisms of nerve and blood vessel wiring. Nature {\bf 436}, 193-200 (2005).
\bibitem{GG05} R.F. Gariano, and T.W. Gardner, Retinal angiogenesis in development and disease. Nature {\bf 438}, 960-966 (2005).
\bibitem{fru07} M. Fruttiger, Development of the retinal vasculature. Angiogenesis {\bf 10}, 77-88 (2007).
\bibitem{lio77} L.A. Liotta, G.M. Saidel, and J. Kleinerman, Diffusion model of tumor vascularization. Bull. Math. Biol. {\bf 39}, 117-128 (1977).
\bibitem{sto91} C. L. Stokes, and D. A. Lauffenburger, Analysis of the roles of microvessel endothelial cell random motility and chemotaxis in angiogenesis. J. Theoret. Biol. {\bf 152}, 377-403 (1991).
\bibitem{cha93}
M.A.J. Chaplain, and A. Stuart, A model mechanism for the chemotactic response of endothelial cells to tumour angiogenesis factor. IMA J. Math. Appl. Med. Biol. {\bf 10}, 149-168 (1993). 

\bibitem{cha95} M.A.J. Chaplain, The mathematical modelling of tumour angiogenesis and invasion. Acta Biotheor. {\bf 43}, 387-402 (1995).

\bibitem{and98} A. R. A. Anderson, and  M. A. J. Chaplain, Continuous and discrete mathematical models of tumor-induced angiogenesis. Bull. Math. Biol. {\bf 60}, 857-900 (1998).

\bibitem{ton01} S. Tong, and F. Yuan, Numerical simulations of angiogenesis in the cornea. 
Microvascular Research {\bf 61}, 14-27 (2001).

\bibitem{lev01} H.A. Levine, S. Pamuk, B.D. Sleeman, and M. Nilsen-Hamilton, Mathematical modeling of the capillary formation and development in tumor angiogenesis: penetration into the stroma. Bull. Math. Biol. {\bf 63}, 801-863 (2001).

\bibitem{pla03} M. J. Plank,  and B. D. Sleeman, Lattice and non-lattice models of tumour angiogenesis. Bull. Math. Biol. {\bf 66}, 1785-1819 (2004).

\bibitem{man04} N.V. Mantzaris, S. Webb, H.G. Othmer, Mathematical modeling of tumor-induced angiogenesis. J. Math. Biol. {\bf 49}, 111-187 (2004).

\bibitem{sun05a}S. Sun, M. F. Wheeler, M. Obeyesekere, and C. W. Patrick Jr., A deterministic model of growth factor-induced angiogenesis. Bull. Math. Biol. {\bf 67}, 313-337 (2005).

\bibitem{sun05}S. Sun, M. F. Wheeler, M. Obeyesekere, and C. W. Patrick Jr., Multiscale angiogenesis modeling using mixed finite element methods. Multiscale Model Simul. {\bf 4},  1137 (2005).

\bibitem{ste06} A. St\'{e}phanou, S. R.  McDougall, A. R. A. Anderson, and M. A. J. Chaplain, Mathematical modelling of the influence of blood rheological properties upon adaptative tumour-induced angiogenesis. Mathematical and Computer Modelling {\bf 44}, 96-123 (2006).

\bibitem{bau07} A.L. Bauer, T.L. Jackson, and Y. Jiang, A cell-based model exhibiting branching and anastomosis during tumor-induced angiogenesis. Biophys. J. {\bf 92}, 3105-3121 (2007).

\bibitem{cap09} V. Capasso, and  D. Morale, Stochastic modelling of tumour-induced angiogenesis. J. Math. Biol. {\bf 58},  219-233 (2009).

\bibitem{jac10}
T. Jackson, and X. Zheng, A cell-based model of endothelial cell migration, proliferation and maturation during corneal angiogenesis. Bulletin of Mathematical Biology {\bf 72}(4), 830-868 (2010).

\bibitem{das10} A. Das, D.A. Lauffenburger, H. Asada, and R. D. Kamm, A hybrid continuum-discrete modelling approach to predict and control angiogenesis: analysis of combinatorial growth factor and matrix effects on vessel-sprouting morphology. Phil. Trans. Roy. Soc. A {\bf 368}, 2937-2960 (2010).

\bibitem{swa11}
K.R. Swanson, R.C. Rockne, J. Claridge, M.A. Chaplain, E.C. Alvord Jr, and A.R.A. Anderson, Quantifying the role of angiogenesis in malignant progression of gliomas: in silico modeling integrates imaging and histology. Cancer Res. {\bf 71}, 7366-7375 (2011).

\bibitem{sci11} M. Scianna, L. Munaron, and L. Preziosi, A multiscale hybrid approach for vasculogenesis and related potential blocking therapies. Prog. Biophys. Mol. Biol. {\bf 106}(2), 450-462 (2011).

\bibitem{sci13} M. Scianna, J. Bell, and L. Preziosi, A review of mathematical models for the formation of vascular networks. J. Theor. Biology {\bf 333}, 174-209 (2013).

\bibitem{cot14}
S.L. Cotter, V. Klika, L. Kimpton, S. Collins, and A. E. P. Heazell, A stochastic model for early placental development. J.R. Soc. Interface {\bf 11}, 20140149 (2014).

\bibitem{dej14} E. Dejana and M.G. Lampugnani, Differential adhesion drives angiogenesis. Nature Cell Biol. {\bf 16}, 305-306 (2014).

\bibitem{ben14}
K. Bentley, C.A. Franco, A. Philippides, R. Blanco, M. Dierkes, V. Gebala, F. Stanchi, M. Jones, I.M. Aspalter, G. Cagna, S. Westr\"om, L. Claesson-Welsh, D. Vestweber, and H. Gerhardt, The role of differential VE-cadherin dynamics in cell rearrangement during angiogenesis. Nat. Cell Biol. {\bf 16}(4), 309-321 (2014).

\bibitem{bon14} L.L. Bonilla, V. Capasso, M. Alvaro, and M. Carretero, Hybrid modeling of tumor-induced angiogenesis. Phys. Rev. E {\bf 90}, 062716 (2014). 

\bibitem{hec15}
T. Heck, M.M. Vaeyens, H. Van Oosterwyck, Computational models of sprouting angiogenesis
and cell migration: towards multiscale mechanochemical models of angiogenesis. Math. Model. Nat.
Phen. {\bf 10}, 108-141 (2015).
\bibitem{ter16} F. Terragni, M. Carretero, V. Capasso, and L.L. Bonilla, Stochastic model of tumour-induced angiogenesis: Ensemble averages and deterministic equations. Phys. Rev. E {\bf 93}, 022413 (2016). 
\bibitem{bon16} L.L. Bonilla, M. Carretero, F. Terragni, and B. Birnir, Soliton driven angiogenesis. Sci. Rep. {\bf 6}, 31296 (2016). doi:10.1038/srep31296.
\bibitem{MS04} N. Manton and P. Sutcliffe,  \emph{Topological solitons} (Cambridge U.P., Cambridge UK 2004).
\bibitem{rem99} M. Remoissenet,  \emph{Waves called solitons: Concepts and experiments}, 3rd ed. (Springer, Berlin 1999).
\bibitem{car16} A. Carpio and G. Duro, Well posedness of an integrodifferential kinetic model of Fokker-Planck type for angiogenesis. Nonlinear Analysis: Real World Applications {\bf 30}, 184-212 (2016). 
\bibitem{BT10} L. L. Bonilla, and  S.W. Teitsworth, \emph{Nonlinear Wave Methods for Charge Transport}  (Wiley-VCH, Weinheim, 2010).
\bibitem{mer97} F. G. Mertens, H. J. Schnitzer, and A. R. Bishop, Hierarchy of equations of motion for nonlinear coherent excitations applied to magnetic vortices. Phys. Rev. B {\bf 56}, 2510-2520 (1997).
\bibitem{san14} B. S\'anchez-Rey, N. R. Quintero, J. Cuevas-Maraver, and M. A. Alejo, Collective coordinates theory for discrete soliton ratchets in the sine-Gordon model. Phys. Rev. E {\bf 90}, 042922 (2014).
\bibitem{gar10} C. W. Gardiner, {\em Stochastic methods. A handbook for the natural and social sciences, 4th ed} (Springer, Berlin 2010).
\bibitem{CDN16} A. Carpio, G. Duro, and M. Negreanu, Constructing solutions for a kinetic model of angiogenesis in annular domains. Preprint arXiv:1612.07389, to appear in Applied Mathematical Modelling. doi:10.1016/j.apm.2016.12.028
\end{thebibliography}
\end{document}